\documentstyle[preprint,aps,epsfig,fancybox]{revtex}
\begin{document}
\preprint{SSF96-09-02}
\draft

\title{Systematic study of Coulomb distortion effects in exclusive
(e,e$'$p) reactions}
\author{V. Van der Sluys, K. Heyde, J. Ryckebusch\   and M. Waroquier\\ 
Department of Subatomic and Radiation Sciences\\
University of Gent \protect\\
 Proeftuinstraat 86 \protect\\
B-9000 Gent, Belgium}

\date{\today}

\maketitle

\begin{abstract}
\noindent A technique to deal with Coulomb electron distortions in the
analysis
of (e,e$'$p) reactions is presented. Thereby, no
approximations are made. The suggested technique relies on a
partial-wave expansion of the electron wave functions and a multipole
decomposition of the electron and nuclear current in momentum space. 
In that way, we
succeed in keeping the computational times within reasonable limits.
This theoretical framework is used to calculate the  quasielastic
(e,e$'$p) reduced
cross sections for proton knockout from the valence shells in
$^{16}$O, $^{40}$Ca, $^{90}$Zr and $^{208}$Pb.
The final-state interaction of the ejected proton with the residual
nucleus is treated within an optical potential model.
The role of electron distortion on the extracted spectroscopic factors
is discussed.
\end{abstract}


\section{Introduction}

For a long time it has been recognized that the
exclusive (e,e$'$N) reaction in the quasielastic (QE) region is a powerful
 tool for studying the single-particle motion inside the
nucleus, and is a testing ground for the different available nuclear models.
One of the principal interests in the exclusive (e,e$'$N) reaction 
is to extract
the nucleon spectral function $P(\vec{p},E)$ from the cross
section. This spectral function can be interpreted as
the joint probability to remove a nucleon
with momentum $\vec{p}$ from the target nucleus and to find 
the residual system
at an excitation energy $E$.
Related to these spectral functions, spectroscopic factors and
occupation numbers are often studied.
They are a measure for the validity
of the independent particle model (IPM). 
The spectroscopic factor $S_{nljm}(E)$ gives the
{\em probability} to reach the single-particle state specified by the
quantumnumbers $nljm$ in the residual
nucleus at an excitation energy $E$. 
The occupation number $N_{nljm}$ 
gives the
number of nucleons in the single-particle state
 $nljm$ in the target nucleus and
involves an integration of the spectroscopic factors 
over the complete excitation energy range \cite{mah85}.
In the IPM the states above (under) the Fermi level are completely
empty (filled) and the total hole (particle) strength is situated
 at a fixed single-particle energy.
The deviation from full (no) occupancy for the
orbits below (above) the
Fermi level is a measure for correlations neglected in this
mean-field approach.

The occupation probabilities in
even-even nuclei have been calculated within several theoretical frameworks.
Most  models go beyond the
mean-field approach and partially account for 
short- and/or long-range nucleon-nucleon correlations
\cite{mah85,vne91,pan84,mut942,mut95,pol95}. 
Occupation probabilities for the
single-particle states 
which considerably deviate from the IPM value were obtained.  Moreover, it is
demonstrated that the single-particle hole strength is fragmented 
over a broad range of energy. 
In particular,
occupation numbers for the proton $3s1/2$ orbit in $^{208}$Pb have
been
 calculated 
varying from $1.42$ \cite{pan84} to $1.66$ \cite{mah91} pointing
towards a strong depletion of this hole state in the ground
state of $^{208}$Pb.
From an experimental point of view, the CERES method \cite{gra92}
 was developed in an attempt  to obtain absolute occupation numbers from
experimental data. The model uses only
relative spectroscopic factors and  allows to account, in an
approximate way, for the strengths at high missing energies, not
accessible for experiment.
With this method, the $3s1/2$ occupation number in $^{208}$Pb is
found to be $1.57(10)$.

Although the advantages of the quasielastic (e,e$'$N) process 
to study spectroscopic
factors are widely recognized, the extraction of these factors from
experiment is still not free of ambiguities. 
For example, depending on the model used in the analysis of the
$^{208}$Pb(e,e$'$p) reaction, the spectroscopic factor for the
transition to the groundstate in $^{207}$Tl ( $3s1/2$
hole) varies from $0.40$ \cite{qui88} to $0.71$ \cite{jin922}.
A reliable  determination
of spectroscopic factors requires an accurate knowledge of the
(e,e$'$N) reaction mechanism (photoabsorption mechanism, final-state
interaction (FSI) of the ejected nucleon with the residual nucleus) 
and the exact treatment
of the Coulomb distortion of the scattered electrons,
especially for heavy nuclei.

In this paper we present results from systematic calculations of 
(e,e$'$p) cross sections for a number of even-even
target nuclei and various 
kinematical conditions
and confront them
with data taken at NIKHEF. The extracted spectroscopic factors
are compared with the corresponding values deduced within other
theoretical approaches \cite{jin922,bof93,udi93,jin91}.
Much  attention is paid to the effect of
electron distortion on the calculated cross section. 
It is pointed out
that, especially
for scattering off heavy nuclei,
 an exact treatment of these effects is highly needed in order
to reproduce the shape of the measured cross sections and, consequently,
to obtain reliable spectroscopic factors. 

This paper is organized as follows. In section II the theoretical
formalism for the (e,e$'$N) reaction  is
outlined. The derivation of the cross section is divided in two
subsections treating the electron and the nuclear aspect of the
(e,e$'$N) reaction. The  technical details are dealt with in
appendix~\ref{appa}.
The numerical details of the adopted approach are discussed in
section III.
The formalism is applied to electro-induced one-proton knockout
reactions from a number of medium-heavy target nuclei in section IV. 
Finally, some
conclusions are drawn in section V.

\section{Formalism}

\subsection{Cross section}

In this paper we describe the process in which an electron 
with four-momentum $k(\epsilon,\vec{k})$ and spin polarization $m_{s_k}$ 
is scattered from a target nucleus at rest with a rest mass $M_A$.
The detected electron is characterized by its four-momentum
$k'(\epsilon',\vec{k}')$ and spin polarization
$m_{s_k'}$.
The energy
transfer to the nucleus $\omega=\epsilon-\epsilon'$ 
is supposed to be sufficient to eject a nucleon $N$ (proton or
neutron) 
with four-momentum
$p_N (E_N,\vec{p}_N)$ and spin projection $m_{s_N}$ out of the target
nucleus leaving the residual nucleus with four-momentum 
$p_B ( E_B,\vec{p}_B)$.
The differential cross section  and the Feynman amplitude
$m_{fi}$ for this process are related as follows:
\begin{eqnarray}
\label{crossgen}
\frac{{\rm d}^4\sigma}{{\rm d}\epsilon'{\rm d}\Omega_e {\rm d}\Omega_N
{\rm d} E_N}
 =\frac{1}{(2\pi)^5}  
 {\epsilon'}^{2} |\vec{p}_N| E_N 
\overline{\sum_{i,f}}
 | m_{fi} | ^2 \; \delta(\omega-S_N-E_x-E_N-E_B+M_N+M_B)\;.
\end{eqnarray}
Throughout this paper we adopt
 natural and unrationalized Gaussian ($\alpha=e^2$) units.
In this
relation $S_N$  stands for the separation energy of a nucleon out of
the target nucleus and $E_x$ denotes the excitation energy
of the residual nucleus. The rest masses of the ejected nucleon and
 the residual nucleus are
given by $M_N$ and $M_B$.
The angles $\Omega_e (\theta_e,\phi_e)$ and $\Omega_N
(\theta_N,\phi_N)$ specify the scattered electron and ejected nucleon
with respect to the chosen reference frame. At this point this
reference frame is not further specified. 
The sum $\overline{\sum_{i,f}}$ implies a summation over all final
states (electron and nuclear) 
and an average over the initial states (electron and nuclear).
We only have to sum over these final
states which satisfy the energy conservation relation. 

In the Born approximation the
transition amplitude $m_{fi}$ can be written in terms of 
matrixelements of the
electron $J_{{\rm el}}^{\mu}$ and nuclear $J_{{\rm nucl}}^{\mu}$
charge-current four-vector in momentum space in
the following way:
\begin{eqnarray}
\label{feynam}
m_{fi}
&=&-\frac{1}{2\pi^2}
\int {\rm d}\vec{q}\; \frac{1}{\omega^2-|\vec{q}\;|^2+i\eta} 
\sum_{\mu} <f_e|J_{{\rm el},\mu}(-\vec{q}\;)|i_e> 
 <f_n|J_{{\rm nucl}}^{\mu}(\vec{q}\;)
|i_n> \;.
\end{eqnarray}
The initial  and final electron states are denoted by $|i_e>$ and
$|f_e>$. The target nucleus and final nuclear state consisting of a residual
nucleus and an ejected nucleon are represented by $|i_n>$ and $|f_n>$.

The Feynman amplitude
$m_{fi}$  can further be rewritten as follows
\begin{eqnarray}
\label{feyncdwba}
\!\!\!\!\!\!\!\!\!\!m_{fi}&=& \frac{1}{2\pi^2}\int{\rm d}\vec{q} 
\left\{
{\frac{1}{|\vec{q}\;|^2}} 
<f_e|\rho_{{\rm el}}(-\vec{q}\;)|i_e> 
{<f_n |\rho_{{\rm nucl}}(\vec{q}\;)|i_n>}
^{\;^{\;^{\;^{\;^{\;^{\;^{\;^{\;}}}}}}}}
_{\;_{\;_{\;_{\;_{\;_{\;_{\;_{\;}}}}}}}} 
\right. \nonumber 
\\
&+&\left. \frac{1}{\omega^2-|\vec{q}\;|^2+i\eta} 
\left[
\sum_{\lambda_q=\pm 1} (-1)^{\lambda_q}
<f_e|J_{{\rm el},\lambda_q}(-\vec{q}\;)|i_e> 
<f_n |J_{{\rm nucl},-\lambda_q}(\vec{q}\;)|i_n> 
\right]
\right\}.
\end{eqnarray}
The spherical components of the electron and nuclear
current operators are taken with
respect to the rotating reference frame  $(x_q,y_q,z_q)$
(Fig.~\ref{kincoul2.eps}~(a)). In this way
the third component of the current operator is directly related to the
charge operator through the charge-current conservation relation.

\subsection{The leptonic part}
\label{seclep}

In this section we  elaborate on the electron matrixelement
$<f_e|J_{{\rm el},\mu}(-\vec{q}\;)|i_e> $ in the expression for the
Feynman amplitude.
The relativistic electron charge-current operator in coordinate space
reads
\begin{eqnarray}
\label{elcur}
\left\{
\begin{array}{l}
J_{\rm el}^0(\vec{r}) = -e \; \hat{\Psi}^{e \;\dagger} (\vec{r}) \; 
\hat{\Psi}^{e}(\vec{r})\;, \\
\vec{J}_{\rm el}(\vec{r}) = -e \; \hat{\Psi}^{e \;\dagger} (\vec{r}) 
\; \vec{\alpha} \;
\hat{\Psi}^{e}(\vec{r}) \;,
\end{array}
\right.
\end{eqnarray}
with $\hat{\Psi}^{e}(\vec{r})$ the electron field operator in
coordinate space. The initial and final electron wave functions are
defined according to
\begin{eqnarray}
<\vec{r}\;|i_e> = \Psi^e_{\vec{k}} (\vec{r}) \;, \\
<\vec{r}\;|f_e> = \Psi^e_{\vec{k}'} (\vec{r}) \;,
\end{eqnarray}
and stand for four-dimensional Dirac spinors.
They are solutions of
the stationary electron Dirac equation:
\begin{eqnarray}
\label{dirac}
(\vec{\alpha}.
(-i\vec{\nabla}) + \beta m_e + V) \; \Psi^e_{\vec{k}m_{s_k}} (\vec{r})
= \epsilon \; \Psi^e_{\vec{k}m_{s_k}} (\vec{r})\;,
\end{eqnarray}
where $m_e$ is the rest mass of the electron and $V$ is the scattering
potential. 
 The additional quantumnumber $m_{s_k}$
uniquely determines the electron wave function.

Dealing with high-energetic electrons the electron mass can
be neglected with respect to its total energy and the Dirac equation
can be written down in the ultrarelativistic limit $(\epsilon=|\vec{k}|)$.
In the Dirac-Pauli  representation for the $\vec{\alpha}$ and $\beta$
matrices and
 in the absence of an external potential $V$
 the solutions of equation~(\ref{dirac}) are given by :
\begin{eqnarray}
\label{freeelwa}
\Psi^{e}_{\vec{k}m_{s_k}}(\vec{r})&=& 
u_e(\vec{k},m_{s_k}) \;e^{i \vec{k}.\vec{r}}
\nonumber \\
&=&
\frac{1}{\sqrt{2}} 
\left(
\begin{array}{c}
\chi^{1/2}_{m_{s_k}}(\Omega_k) \\ 
\frac{\vec{\sigma}.\vec{k}}{|\vec{k}|}\chi^{1/2}_{m_{s_k}}(\Omega_k) 
\end{array}
\right)
\;e^{i \vec{k}.\vec{r}}\;.
\end{eqnarray}
The spinors $\chi^{1/2}_{m_{s_k}}(\Omega_k)$ can be expressed in terms
of the Pauli spinors and the matrixelements of the Wigner 
${\cal D}^{\frac{1}{2}}$-matrix, i.e.,
\begin{eqnarray}
\chi^{1/2}_{m_{s_k}}(\Omega_k) = \sum_{m_s} \chi^{1/2}_{m_{s}}(\sigma){\cal
D}^{\frac{1}{2}}_{m_s m_{s_k}}(\varphi_k,\theta_k,0) \;.
\end{eqnarray}
The angles $\Omega_k=(\theta_k,\varphi_k)$ specify the momentum
$\vec{k}$ with respect to the chosen reference frame
($x,y,z$). 
The Wigner ${\cal D}^{\frac{1}{2}} (R_k)$ matrix represents 
the rotation of the reference frame ($x,y,z$) over the
Euler angles $R_k=(\varphi_k,\theta_k,0)$
in the basis spanned by the eigenvectors of the operators
$\hat{S}^2$ and $\hat{S}_z$. 

Assuming a central potential $V=V(r)$, the electron wave functions are
evaluated by a phase shift analysis based on a partial-wave expansion.
Indeed, 
the Dirac Hamiltonian $(\hat{H}=\vec{\alpha}.\vec{k} + V(r))$ commutes with
the angular momentum operators $\hat{J}^2$ and $\hat{J}_z$ and
with the operator $\hat{K}=\beta\;\{\vec{\sigma}.\vec{L} + 1\}$ but
not with the
orbital momentum operator $\hat{L}^2$.
As such, we derived a complete set of operators with common
eigenfunctions represented by 
$\tilde{\Psi}^{\epsilon}_{\kappa jm}(\vec{r})$:
\begin{eqnarray}
\label{eigenrel}
\left\{
\begin{array}{l}
\hat{H} \; \tilde{\Psi}^{\epsilon}_{\kappa jm}(\vec{r})\;=\;
\epsilon\;\tilde{\Psi}^{\epsilon}_{\kappa jm}(\vec{r}) \; ,
\\
\hat{J}^2 \; \tilde{\Psi}^{\epsilon}_{\kappa jm}(\vec{r})\;=\;
j(j+1)\;\tilde{\Psi}^{\epsilon}_{\kappa jm}(\vec{r}) \; ,
\\
\hat{J}_z \;\tilde{\Psi}^{\epsilon}_{\kappa jm}(\vec{r})\;=\;
m\;\tilde{\Psi}^{\epsilon}_{\kappa jm}(\vec{r}) \; ,
\\
\hat{K}\;\tilde{\Psi}^{\epsilon}_{\kappa jm}(\vec{r})\;=\; -\kappa
\;\tilde{\Psi}^{\epsilon}_{\kappa jm}(\vec{r})\;.
\end{array}
\right.
\end{eqnarray}
We can construct the
partial waves $\tilde{\Psi}^{\epsilon}_{\kappa jm}(\vec{r})$ as follows:
\begin{eqnarray}
\label{parwav}
\tilde{\Psi}^{\epsilon}_{\kappa
jm}(\vec{r})&=&\Psi^{\epsilon}_{ljm}(\vec{r})=
\left(
\begin{array}{c}
\frac{G^{\epsilon}_{lj}(r)}{r} {\cal Y} ^{jm}_{l 1/2} (\Omega_r,\sigma) 
\\
i\;\frac{F^{\epsilon}_{lj}(r)}{r} 
{\cal Y} ^{jm}_{\overline{l} 1/2} (\Omega_r,\sigma)
\end{array}
\right)
\\ \nonumber &&{\rm with}\quad
\left\{
\begin{array}{ll}
l=j-1/2 &\qquad{\rm if}\quad \kappa=-(j+1/2) \; ,
\\
l=j+1/2 &\qquad{\rm if}\quad \kappa=j+1/2 \;.
\end{array}
\right. 
\end{eqnarray}
We introduce the common notation $\overline{l}$ 
\begin{eqnarray}
\left\{
\begin{array}{c}
l= j + \frac{1}{2} \;\; \Rightarrow \;\;  \overline{l}=j-\frac{1}{2}\;,
\\
l= j - \frac{1}{2} \;\; \Rightarrow \;\;  \overline{l}=j+\frac{1}{2}\;.
\end{array}
\right .
\end{eqnarray} 
The spherical spin-orbit eigenspinor 
${\cal Y} ^{jm}_{l 1/2} (\Omega_r,\sigma)$
is defined in the following way
\begin{eqnarray}
{\cal Y} ^{jm}_{l 1/2} (\Omega_r,\sigma)
= \sum_{m_l m_s} < l m_l 1/2 m_s | j m> {\rm Y}_{lm_l} (\Omega_r) 
\chi^{1/2} _{m_s}(\sigma)\;.
\end{eqnarray}
Each partial wave (\ref{parwav}) can be easily proved 
to satisfy the eigenvalue equations
(\ref{eigenrel}) under the condition that
 the radial electron wave functions $G^{\epsilon}_{lj}(r)$ and 
$F^{\epsilon}_{lj}(r)$
are
solutions of the following second-order differential equations:
\begin{eqnarray}
\label{second}
\left\{ \begin{array}{l}
\frac{{\rm d}^{2} }{{\rm d}r^{2}}\;G^{\epsilon}_{lj}(r) +
\frac{{\rm d}V(r)/{\rm d}r}{E-V(r)} 
\frac{{\rm d}}{{\rm d}r}\;G^{\epsilon}_{lj}(r)
\\ \qquad+ \left[(E-V(r))^{2} -
\frac{\kappa(\kappa+1)}{r^{2}} +
\frac{\kappa}{r}
\frac{{\rm d}V(r)/{\rm d}r}{E-V(r)} \right]
\;G^{\epsilon}_{lj}(r) = 0 \;,\\ 
\frac{{\rm d}^{2} }{{\rm d}r^{2}}\;F^{\epsilon}_{lj}(r) +
\frac{{\rm d}V(r)/{\rm d}r}{E-V(r)} 
\frac{{\rm d}}{{\rm d}r}\;F^{\epsilon}_{lj}(r)
\\ \qquad+ \left[(E-V(r))^{2} -
\frac{\kappa(\kappa-1)}{r^{2}} -
\frac{\kappa}{r}
\frac{{\rm d}V(r)/{\rm d}r}{E-V(r)} 
\right]
\;F^{\epsilon}_{lj}(r) = 0 \;.
\end{array} 
\right.
\end{eqnarray}
For each partial wave $lj$  the second-order
differential equation for $G^{\epsilon}_{lj}(r)$ 
has to be solved numerically.
For the regular solutions one imposes the following boundary conditions:
\begin{eqnarray}
\lim_{r \rightarrow 0} G^{\epsilon}_{lj}(r) &=& 0 \;,
\nonumber
\\
\lim_{r \rightarrow 0} \frac{\rm d}{{\rm d}r}\;G^{\epsilon}_{lj}(r) &=&
0  \quad{\rm for} \quad l>0\;,
\end{eqnarray}
and one obtains the
corresponding solution for $F^{\epsilon}_{\overline{l}j}(r)$ through
the relation:
\begin{eqnarray}
\label{corres}
G^{\epsilon}_{lj}(r)\;=\;(l-\overline{l})F^{\epsilon}_{\overline{l}j}(r)\;.
\end{eqnarray}

The asymptotic behaviour of the  radial electron wave functions for
Coulomb potential scattering are given by ($k \equiv |\vec{k}|$)
\begin{eqnarray}
\label{asymgf2}
\lim_{r \rightarrow \infty} G^{\epsilon}_{lj}(r)&=& 
(\overline{l}-l)
\;\frac{\sin(kr -l\pi/2 + \delta^{{\rm e},\epsilon\;({\rm tot})}_{lj}
 - \eta \ln(2kr))}{k}\;,
\\
\lim_{r \rightarrow \infty} F^{\epsilon}_{lj}(r)
&=&-\;
\frac{\sin(kr -\overline{l}\pi/2 + 
\delta^{{\rm e},\epsilon\;({\rm tot})}_{\overline{l}j} 
- \eta \ln(2kr))}{k}\;.
\end{eqnarray}
The phase shift $\delta^{{\rm e},\epsilon\;({\rm tot})}_{lj}$ 
reflects the influence
of the scattering potential $V$. 
It consists of two parts, i.e., 
the Coulomb phase shift $\sigma_l^{\rm e}$ and an
additional phase shift $\delta^{{\rm e},\epsilon}_{lj}$. For a Coulomb
potential generated by the $Z$ protons in the nucleus, the Coulomb phase
shift is defined according to ($\eta=-Ze^2$): 
\begin{eqnarray}
\sigma_{l}^{\rm e}=\arg \Gamma (l+1+i\eta)\;.
\end{eqnarray}
Due to the fact that the scattering potential $V$ is spin-independent,
one can easily verify that the total phase shift is
$l$-independent, i.e.,
\begin{eqnarray}
\label{phassh}
\delta^{{\rm e},\epsilon\;({\rm tot})}_{j}=
\delta^{{\rm e},\epsilon\;({\rm tot})}_{lj}=
\delta^{{\rm e},\epsilon\;({\rm tot})}_{\overline{l}j}.
\end{eqnarray}

Finally, the electron wave function 
$\Psi_{\vec{k}m_{s_k}}^{e(\pm)}(\vec{r})$
is expanded in terms of these partial waves
$\Psi^{\epsilon}_{ljm}(\vec{r})$:
\begin{eqnarray}
\label{pardecom}
\Psi_{\vec{k}m_{s_k}}^{e(\pm)}(\vec{r}) = 
\sum_{ljm} a_{ljm}^{\epsilon m_{s_k}\;(\pm)}
\Psi^{\epsilon}_{ljm}(\vec{r}) \;.
\end{eqnarray}
The initial and final electron wave functions have to satisfy the
outgoing ($+$) respectively incoming ($-$) boundary conditions.
Knowing the asymptotic behaviour of the radial electron wave
functions,  the coefficients
$a_{ljm}^{\epsilon m_{s_k}(\pm)}$ are fixed by
\begin{eqnarray}
\label{coef}
\! a_{ljm}^{\epsilon m_{s_k}(\pm)}= \sum_{m_s m_l} {\cal
D}^{\frac{1}{2}}_{m_s m_{s_k}}(R_k) \frac{4\pi}{\sqrt{2}}
i^l e^{\pm \;i\delta^{{\rm e},\epsilon\;({\rm tot})}_{j}} 
(\overline{l} -l)
{\rm Y}^{\ast}_{lm_l}(\Omega_k) <l m_l 1/2 m_s | j m> .
\end{eqnarray}

At this point only the scattering potential $V$ remains to be
specified. In general the central Coulomb scattering
potential generated by $Z$ protons is
 given by
\begin{eqnarray}
V(r)\;&=&\;-4\pi Z\alpha \frac{1}{r}
\int_{0}^{r} \rho(r'){r'}^2 {\rm d} r'
-4\pi Z\alpha \int_{r}^{\infty} \rho(r')r' {\rm d} r'\;, 
\end{eqnarray}
with $\rho(r)$  the nuclear
charge density normalized according to
 $4\pi\int_{0}^{\infty} \rho(r)r^2 {\rm d}r = 1$.
In the forthcoming discussion we have taken this charge density
to 
correspond with a
homogeneous spherical charge distribution of $Z$ protons
within the nuclear radius $R$.

By switching off the scattering potential $V$ one can 
easily verify that 
the solution (\ref{pardecom})
coincides with the free electron wave function (\ref{freeelwa}) since
the differential equations (\ref{second}) reduce to the differential
equations for the spherical Bessel functions. 
In this way a sensitive testing case for our numerical approach is found.

We want to stress that 
the problem of Coulomb distortion of the initial and
final electron in the electron scattering process is solved to all
orders. 
Earlier work
in this field by Boffi {\em et al.} \cite{bof93}
 handled the electron distortion in an approximate way
through
 a high-energy expansion of the
electron wave functions combined with an expansion in powers of
$Z\alpha$. The DWEEPY code \cite{bof93} used in the analysis 
of the NIKHEF data
adopts this approximate treatment of electron distortion.
To lowest order in $Z\alpha$ it was proved that electron
distortion effects could be approximated by an effective momentum
approach (EMA). 
This means that the plane wave in eq.~(\ref{freeelwa}) has to
be replaced by
\begin{eqnarray}
e^{i \vec{k}.\vec{r}} \longrightarrow \frac{k^{{\rm eff}}}{k}
 e^{i\vec{k}^{{\rm eff}}.\vec{r}}\;,
\end{eqnarray}
with
\begin{eqnarray}
\label{keff}
\vec{k}^{{\rm eff}}=(k+\frac{3Z\alpha}{2R})\vec{e}_k \;.
\end{eqnarray}
Clearly this approach is very easy to handle and  worth
comparing with the
complete distorted wave approach so that its degree of accuracy can be
estimated.

\subsection{The nuclear part}

In a previous paper \cite{vsl96}, 
we have shown that at low values of the missing
momentum, meson-exchange currents (MEC) and long-range effects only
slightly affect the calculated (e,e$'$p) cross section. 
As we will restrict ourselves to QE (e,e$'$p) reactions 
at low missing momenta only
the one-body part of the nuclear four-current is retained. Hereby we
adopt the operator as dictated 
in the non-relativistic impulse approximation:
\begin{eqnarray}
\label{onecur}
\rho_{{\rm nucl}}(\vec{r}) &=& \sum_{i=1...A} e G_E^{i}(\vec{r},\omega)
\delta(\vec{r}-\vec{r}_i) \;,
\\
\vec{J}_{{\rm nucl}}(\vec{r}) &=& \sum_{i=1...A} 
\left\{ 
\frac{e G_E^{i}(\vec{r},\omega)}{i2M_i}
\left(\vec{\nabla}_i \delta(\vec{r}-\vec{r}_i) +
\delta(\vec{r}-\vec{r}_i)\vec{\nabla}_i \right) 
\right.
\nonumber\\&&\qquad\quad
\left.
+ \frac{e G_M^{i}(\vec{r},\omega) }{2M_i}
\delta(\vec{r}-\vec{r}_i)\vec{\nabla}\times\vec{\sigma}_i
\right\}\;.
\nonumber
\end{eqnarray}
This nuclear charge-current four-vector refers to
 $A$ non-interacting point-like
nucleons with mass $M_i$.
To correct for the  finite
extent of the nucleons, the Sachs electromagnetic formfactors $G_E$
and $G_M$ are introduced.

As for the electron wave functions, 
the final nuclear wave function is determined through a phase shift
analysis after an expansion
 in partial waves.
The final nuclear state is taken to be a linear combination of
one particle-one hole excitations $|C;\omega JM>$ out of  the A-particle
groundstate $|i_n>$ with 
$C \equiv \{h,p\}$. The hole state $h$ is characterized by the
quantumnumbers $n_h,l_h,j_h$ and energy $\epsilon_h$. 
The continuum particle
state  is specified by the quantumnumbers $p=(l,j)$ and the energy 
$\epsilon_p=E_N - M_N$. The isospin
nature of the particle-hole state is denoted by $t_q$.
The particle-hole state in the coupled scheme is defined according to
\begin{eqnarray}
|C ;\omega JM>&=& \sum_{m_h m} < j_h -m_h j m | JM >
(-1)^{j_h-m_h} |ph^{-1}(\omega)>\;,
\end{eqnarray}
with the uncoupled particle-hole state defined as follows
\begin{eqnarray}
|p h^{-1} (\omega)> = c^{+}_p(\epsilon_p) c_h  |i_n> \;,
\end{eqnarray}
and $\omega=\epsilon_p-\epsilon_h$.
The operators $c^{+}$ and $c$ denote single-particle creation and
annihilation operators.
The radial wave functions for the bound
hole states are solutions of the Schr\"{o}dinger equation with a
Hartree-Fock potential generated with
an effective interaction of the Skyrme type (SkE2) \cite{war87}. 
The continuum particle states are evaluated within
an optical potential model (OPM) \cite{bof93}.
The physical radial wave functions are regular in the origin and behave
asymptotically $(r \rightarrow \infty)$ according to 
\begin{eqnarray}
\label{boundnucl}
\left\{
\begin{array}{ll}
\phi_{p}(r) \stackrel{r \rightarrow \infty}{\longrightarrow}
\sqrt{\frac{2\mu_N}{\pi k_p}} \frac{\sin(k_p r -  l\pi/2 -\eta \ln 2k_pr +
\delta_{lj}^{{\rm n},\epsilon_p({\rm tot})})}{r} 
&  \epsilon_p > 0\;, \\\\
\phi_{h}(r) \stackrel{r \rightarrow \infty}{\longrightarrow} 0
&  \epsilon_h < 0 
\end{array}
\right.
\end{eqnarray}
where $\eta$ and the momentum $k_p\equiv |\vec{k}_p|$ stand for
\begin{eqnarray}
\begin{array}{ll}
k_p^2 = 2\mu_N\epsilon_p & \mbox{\rm with}\; \mu_N=M_N (A-1)/A
\;\;\mbox{\rm the reduced mass of the nucleon}\;, \\ 
\eta= \frac{(Z-1)\alpha \mu_N}{k_p}\;. &
\end{array}
\end{eqnarray}
The complex phase shifts caused by the nuclear and Coulomb part of the
optical potential are denoted by $\delta^{{\rm n},\epsilon_p}_{lj}$
and $\sigma^{\rm n}_{l}$
  ($\delta_{lj}^{{\rm n},\epsilon_p {\rm (tot)}}=
\delta^{{\rm n},\epsilon_p}_{lj}+\sigma_{l}^{\rm n}$).

Given the asymptotic behaviour for the radial single-particle 
wave functions and
imposing that 
the ejected nucleon wave function
 satisfies the incoming boundary conditions, the final
nuclear state $|f_n>$ is given by
\begin{eqnarray}
\label{nuclwav}
\mid f_n>
& = & \sum_{ljmm_{l}} \sum_{JM} 4\pi
i^{l}\sqrt{\frac{\pi}{2\mu_{N}k_{p}}} < j_{h}m_{h} j m \mid J M >
< l m_{l} \frac{1}{2} m_{s} \mid j m >
 \nonumber\\&&\qquad\qquad\quad\times 
e^{-i\delta_{lj}^{{\rm n},\epsilon_p {\rm (tot)}}}
{\rm Y}^{\ast}_{lm_{l}}(\Omega_{N})\mid (l_h j_h, lj);\omega JM >\;.
\end{eqnarray}
In order to derive this expression
the target nucleus is considered to be a spherical nucleus in the
$J^{\pi}=0^+$ groundstate. 
In addition, the residual nucleus is described by a pure hole state $h$
with respect
to this  target nucleus groundstate.

\subsection{The Feynman amplitude}

As the initial and final electron wave function and the final nuclear
state are expanded in partial waves, it is common to
decompose the electron and nuclear charge-current operators 
in the Coulomb, Electric and
Magnetic multipole operators of rank $JM$ ($q=|\vec{q}\;|$):
\begin{eqnarray}
\label{tranop}
T^{{\rm el}}_{JM}(q) &=& \frac{1}{q} 
\int {\rm d}\vec{r} \; \vec{\nabla}\times
\left[
j_{J}(qr)\vec{\cal Y}^M_{J(J,1)}(\Omega_r)
\right].\vec{J}(\vec{r}) \;,
\nonumber\\
T^{{\rm mag}}_{JM}(q)&=&\int {\rm d}\vec{r}\;
j_{J}(qr)\vec{\cal Y}^M_{J(J,1)}(\Omega_r).\vec{J}(\vec{r}) \;,
\nonumber\\
M^{{\rm coul}}_{JM}(q)&=&\int {\rm d}\vec{r}
\;j_{J}(qr){\rm Y}_{JM}(\Omega_r)\rho(\vec{r}) \;,
\end{eqnarray}
with the vector spherical harmonics defined according to
\begin{eqnarray}
\vec{\cal Y}^M_{J(L,1)}(\Omega)=\sum_{\lambda M_L}<L M_L 1 \lambda | J
M > {\rm Y}_{LM_L}(\Omega) \vec{e}_{\lambda}\;,
\end{eqnarray}
and $\vec{e}_{\lambda} (\lambda=0,\pm1)$ 
the standard spherical unit vectors
corresponding with the unit vectors ($\vec{e}_x,\vec{e}_y,\vec{e}_z$)
in the $(x,y,z)$ reference frame (Fig.~\ref{kincoul2.eps}~(a)).

Accordingly, in momentum space 
the charge-current operators can be
written as:
\begin{eqnarray}
\label{muldecom}
&&\rho(\vec{q}\;)= 4\pi \sum_{JM} i^J {\rm Y}^{\ast}_{JM} (\Omega_q)
M^{{\rm coul}}_{JM} (q)\;,
\nonumber\\
&&J_{\lambda_q}(\vec{q}\;) = -\sqrt{2\pi} \sum_{J\geq 1,M} i^J \hat{J}
\left[
T^{{\rm el}}_{JM}(q) +\lambda T^{{\rm mag}}_{JM}(q)
\right] {\cal D}^J_{M\lambda}(R_q)
\end{eqnarray}
with $\hat{J}=\sqrt{2J+1}$ and the Euler angles 
$R_q=(\phi_q,\theta_q,-\phi_q)$ defined in Fig.~\ref{kincoul2.eps}~(a).

It is well-known that
when neglecting electron distortion effects, the
differential (e,e$'$N) cross section can be written in terms of
four structure functions containing all the nuclear information. In such
a Distorted Wave Born Approximation (DWBA) approach  
each structure function is
multiplied with an analytical factor containing  the leptonic information.
This is no longer valid  in the Coulomb Distorted 
Wave Born Approximation (CDWBA)
approach as the electron part can no longer be
separated from the nuclear part. 
Consequently, when accounting for Coulomb distortion effects 
one has to perform a
multipole expansion for both the electron and nuclear charge-current
operators.

Combining equations~(\ref{feyncdwba}) and (\ref{muldecom})
 and applying some basic properties of the Wigner ${\cal D}^J
(R_q)$ matrices
 the Feynman amplitude $m_{fi}$ reads as 
($q_{\mu}q^{\mu}\equiv\omega^2-q^2$)
\begin{eqnarray}
\label{fincdwba}
m_{fi}&=& \sum_{LM_{L}} (-1)^{M_{L}}\frac{(4\pi)^{3}}{(2\pi)^3}
 \int_{0}^{\infty} {\rm d}q 
\left[
<f_e|M^{{\rm e},{\rm coul}}_{LM_{L}}(q)|i_e>
<f_n|M^{{\rm n},{\rm coul}}_{L-M_{L}}(q)|i_n> 
\right.\nonumber\\
&&
\left.\qquad\qquad\quad
-\frac{q^{2}}{q_{\mu}q^{\mu}+i\eta}
\left\{ 
<f_e|T^{{\rm e},{\rm mag}}_{LM_{L}}(q)|i_e>
<f_n|T^{{\rm n},{\rm mag}}_{L-M_{L}}(q)|i_n> 
\right.
\right.\nonumber\\
&&
\left.
\left.\qquad\qquad\quad\;\;
+
<f_e|T^{{\rm e},{\rm el}}_{LM_{L}}(q)|i_e>
<f_n|T^{{\rm n},{\rm el}}_{L-M_{L}}(q)|i_n>
\right\}
\right]\;.
\end{eqnarray} 
The superscript ${\rm e}$ and ${\rm n}$ refer to  the electron and the
nuclear multipole operators. 
We have deliberately chosen to work out the leptonic 
and nuclear matrixelements in
momentum space. Earlier electron distortion
calculations by Jin {\em et al.} \cite{jin922} and Ud\'{\i}as {\em et
al.} \cite{udi93}
evaluate the transition
matrixelements in coordinate space. In order to make their
calculations feasible the nucleon formfactors are evaluated  at the
asymptotic value $\vec{q}=\vec{k}-\vec{k}'$. 
The major advantage of 
our approach is  that the momentum-dependence of the nucleon formfactors can 
 be handled
exactly.

Here we will solely calculate
the unpolarized (e,e$'$N) cross section (\ref{crossgen})
so we need to evaluate
\begin{eqnarray}
\label{cdwba1}
\overline{\sum_{i,f}} | m_{fi} | ^2 = 
\frac{1}{2}\sum_{m_{s_k} m_{s_{k'}}} 
\sum_{m_B m_{s_N}}| m_{fi} | ^2 \;.
\end{eqnarray}
 The summation over the initial and final states 
involves a 
sum over the initial and final electron polarizations and a sum over
the polarizations of the recoiling nucleus and the ejected nucleon. 
In appendix~\ref{appa} this Feynman amplitude is further worked out.
Summarizing from appendix~\ref{appa}, one can state
 that the calculation of the (e,e$'$N) cross section 
 is reduced to the evaluation
of a large number of leptonic radial integrals
${\cal R}_{Lj_1j_2}(\epsilon,\epsilon';q)$
and a set of reduced transition matrixelements
${\cal L} (C;q\omega J)$ containing all nuclear information.
We stress that the technique developed here can be easily
extended to polarization processes.

\section{Numerical procedure}

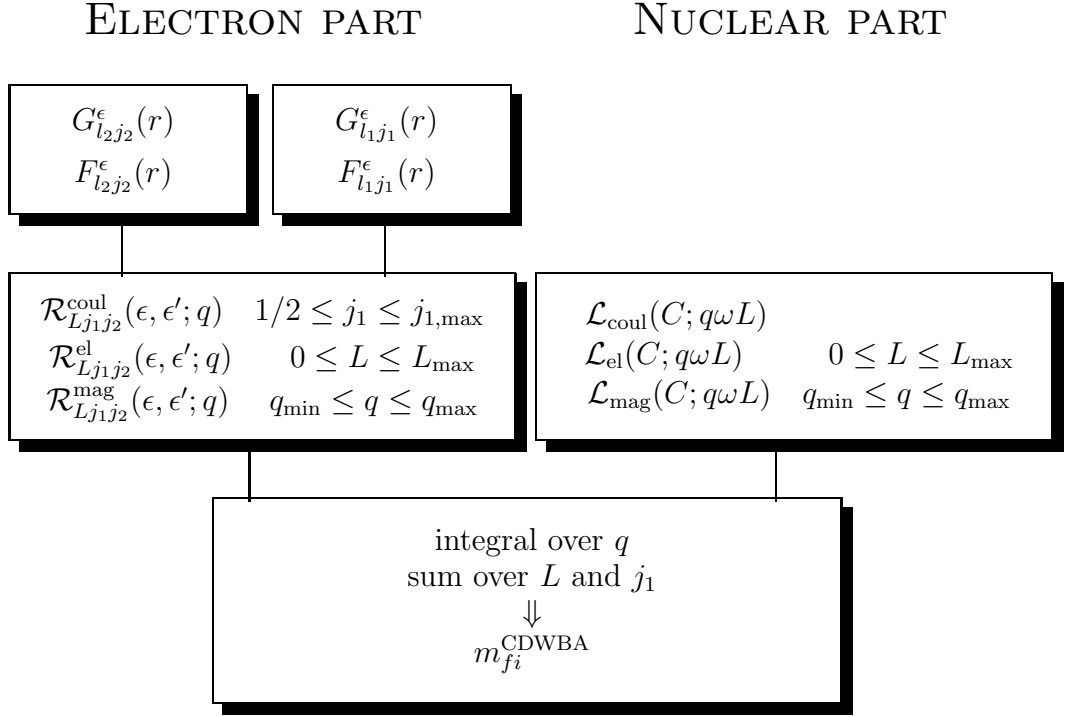
\begin{figure}
\newcounter{cms}
\setlength{\unitlength}{1mm}
\centering
\begin{picture}(145,110)
\put(5,90){\parbox[b][1.5cm][s]{6.5cm}{\begin{center}
{\Large\sc Electron part}
\end{center}}
}
\put(5,70){\shadowbox{\parbox[b][1.5cm][s]{2.8cm}{\begin{center}
{\shortstack{ $G_{l_2j_2}^{\epsilon}(r)$\\ \\
 $F_{l_2j_2}^{\epsilon}(r)$}}
\end{center}}
}}
\put(40,70){\shadowbox{\parbox[b][1.5cm][s]{2.8cm}{\begin{center}
{\shortstack{ $G_{l_1j_1}^{\epsilon}(r)$\\ \\
 $F_{l_1j_1}^{\epsilon}(r)$}}
\end{center}}
}}
\put(20,63.2){\line(0,1){6.8}}
\put(55,63.2){\line(0,1){6.8}}
\put(5,40)
{\shadowbox{\parbox[b][2cm][s]{6.5cm}
{\begin{center}{\shortstack
{ ${\cal R}^{\rm coul}_{Lj_1j_2}(\epsilon,\epsilon';q)
\quad 1/2 \le j_1 \le j_{1,{\rm max}}$
\\
 ${\cal R}^{\rm el}_{Lj_1j_2}(\epsilon,\epsilon';q)
\quad\;\;\; 0 \le L \le L_{\rm max}$
\\
 ${\cal R}^{\rm mag}_{Lj_1j_2}(\epsilon,\epsilon';q)\quad\; q_{\rm
min} \le q \le q_{\rm max}$}}
\end{center}
}}}
\put(75,90)
{ \parbox[b][1.5cm][s]{6.5cm} 
{\begin{center} 
{\Large\sc Nuclear part}
\end{center}
}}
\put(75,40)
{\shadowbox{ \parbox[b][2cm][s]{6.5cm} 
{\begin{center} {\shortstack{
 ${\cal L}_{\rm coul}(C;q\omega L)$
$\quad\;\; \qquad\qquad\qquad$\\
 ${\cal L}_{\rm el}(C;q\omega L)$ 
$\quad\;\;\;\;\; 0 \le L \le L_{\rm max}$
\\
 ${\cal L}_{\rm mag}(C;q\omega L)\quad q_{\rm min} \le q \le q_{\rm
max}$
}}
\end{center}
}}}
\put(37,33.3){\line(0,1){6.7}}
\put(107,33.3){\line(0,1){6.7}}
\put(32,5){\shadowbox{ \parbox[b][2.5cm][s]{8cm}
{\begin{center}{\shortstack{
 integral over $q$ \\
 sum over $L$ and $j_1$\\
$\Downarrow$\\
 $m_{fi}^{\rm CDWBA}$}}
\end{center}
}}}
\end{picture}
\caption{Schematic representation of the CDWBA approach for the
exclusive (e,e$'$p) cross section.}
\label{schemacdwba}
\end{figure}

In order to derive the exclusive (e,e$'$p) cross section (\ref{crossgen})
 in the CDWBA we need to
evaluate the Feynman amplitude $m_{fi}$ discussed in
the previous section and appendix~\ref{appa}. The numerical procedure is
 schematically sketched in Fig.~\ref{schemacdwba}. 
From a numerical point of view the evaluation of this transition
amplitude is cumbersome as it involves  an integration over the complete
$q$ range and two infinite sums, i.e., the
sum over the  different multipolarities $L$ in the multipole
expansion of the leptonic and hadronic current and the 
sum over the angular momentum  $j_1$ originating from
 the partial-wave expansion of the scattered electron state.
Angular momentum selection rules make sure that the other
summations in the equations~(\ref{electroncdwba}) and
(\ref{nuclearcdwba}) have a finite range for  fixed values of $j_1$
and $L$.

When accounting for electron distortion effects, the integrandum in the
integral
 over $q$
peaks at the effective momentum transfer
$q^{\rm eff}=|\vec{k}^{\rm eff}-\vec{k}'^{\rm eff}|$.
As the EMA is only an approximation of electron distortion effects 
the integrandum is spread around this value and
 the integration in $q$-space has to be  performed in an interval 
$[q_{\rm min}, q_{\max}]$
around $q^{\rm eff}$. It is worth noting that in the absence of electron
distortion effects, the integral over $q$
vanishes and the standard DWBA expressions are retained. The
integrandum then reduces to a $\delta$-function representing the momentum
conservation relation $q=|\vec{k}-\vec{k}'|$.

The finite extent of the nucleus puts a constraint on the number of
multipolarities $L$  which have to be retained in the multipole
expansion of the nuclear current given in eq.~(\ref{muldecom}).
In the calculations we systematically observe convergence when including
multipolarities up to
 $L_{\rm max}\approx 2 q R$  where
$R$ denotes the radius of the considered target nucleus.
 In a similar way the number of electron partial waves which
contribute to the (e,e$'$p) cross section, is restricted by an upper limit
$j_{1,{\rm max}}$. 
It can be easily  verified that the electron partial waves 
$G_{lj}^{\epsilon}(r)$ and $F_{lj}^{\epsilon}(r)$ corresponding with
large values for $lj$ are
negligible  for  values of
$r$ within the nucleus range.
For that reason, these electrons can cause no nuclear 
transitions. Consequently, to a required accuracy, only a
finite number of the electron partial waves contributes  to the
electron scattering cross section.
The number of electron partial waves actually contributing to the
cross section depends also on the electron energy. 
The higher the electron energy the more partial waves will be required.
The numerical evaluation of the (e,e$'$p) cross sections 
is getting  complicated
due to the large number  of electron partial waves to consider.
This is a result of the long-range character of the Coulomb interaction. The
limit $Z\rightarrow 0$ (equivalent with turning off the electron
distortions) can be considered as
 a severe test of the accuracy of the numerical techniques
 and a convergence test for the  electron
partial waves. 
For $Z=0$, the electron
wave functions reduce to  plane waves. Accordingly, the DWBA cross
section  should be retained. 
As will be demonstrated in the forthcoming sections, 
our code has been checked to comply with this requirement.

Another important feature of our CDWBA approach is 
that the radial integrals ${\cal R}_{Lj_1j_2}$ (\ref{radint1}),
which are the heart of our numerical procedure, do not  depend on the
scattering angles $\theta_e$ and $\theta_p$. Consequently, our
numerical procedure is optimized for 
calculating the (e,e$'$p) cross section for these specific kinematical
conditions
 where the electron and proton scattering
 angles are
varied and the other electron characteristics are kept fixed. 
The complete missing momentum range of the (e,e$'$p) cross section 
 for proton knockout from the different hole states can 
then be calculated  with a stored set of
radial integrals.

\section{Results}

Up to now, most of the high-resolution  (e,e$'$p) experiments 
performed at NIKHEF 
(Amsterdam), Saclay, Mainz and MIT-Bates have been carried out by using
either parallel or constant $\vec{q}-\omega$ kinematics. 
Both 
correspond with in-plane experiments: the ejected proton is
detected in the scattering plane spanned by the initial and final
electron.
In {\em parallel kinematics} the proton is detected in the direction of the
momentum transfer.
By varying the incoming $\epsilon$ and outgoing $\epsilon'$
 electron energies or/and the
scattering angle $\theta_e$,
 different values for the momentum transfer $\vec{q}=\vec{k}-\vec{k}'$
and consequently
 the missing momentum $\vec{p}_m=\vec{p}_p -\vec{q}$
are reached. 
 For {\em constant $\vec{q}-\omega$ kinematics} the
energy-momentum transfer is kept fixed 
and the proton angular distribution is measured. The missing momentum
is defined positive when the ejected proton lies in the half-plane of the
initial electron momentum and bordered by
the momentum transfer. In the other half-plane  the missing momentum
is negative.

Most of the experimental data are presented in terms of the
reduced cross section  extracted from the measured cross section in the
following way ($p_m=|\vec{p}_m|$, $p_p=|\vec{p}_p|$)
\begin{eqnarray}
\label{extheo}
\rho_m(p_m,E_x)=
\frac{1}{ p_p E_p \sigma_{ep} }
\frac{{\rm d}^4\sigma}{{\rm d}\epsilon'{\rm d}
\Omega_e {\rm d}\Omega_p {\rm d} E_p} 
\end{eqnarray} 
with $\sigma_{ep}$ the off-shell electron-proton cross section.
We stress that only in the plane-wave impulse approximation (PWIA)
 the reduced cross section
coincides with
 the nucleon spectral function $P(p_m,E_x)$, i.e., 
the probability to eject a
nucleon with momentum $p_m$ from the target nucleus  while leaving the
residual nucleus at an excitation energy $E_x$. 
As soon as the FSI, electron distortion and many-body nuclear currents
effects come into play this quantity can no longer be interpreted as 
the nucleon spectral function.
In comparing our (e,e$'$p) results with the available data we have divided
the calculated cross sections with the $cc1$ prescription \cite{for83}
 for $\sigma_{ep}$.
 The same procedure was applied to the
experimental  cross sections
presented in this paper.
Moreover, 
the calculated  curves are scaled with a spectroscopic factor which
accounts for
the fragmentation of the single-particle strength.

The results of our model calculation are compared with the predictions
from three other
model calculations. Firstly, we confront our results with the
non-relativistic CDWBA model of Boffi {\em et al.}\cite{bof93}. This model is 
at the basis of the DWEEPY code often used in the analysis of the NIKHEF
(e,e$'$p) data.
In the latter model the FSI is treated in a non-relativistic optical
potential calculation similar to ours. In contrast with our
model,
the bound state wave functions are calculated in a
Wood-Saxon well. The rms radius of the bound state wave function is fitted to reproduce the
shape of the measured reduced cross section and the well depth is
adjusted to reproduce the experimentally observed separation energy.
In our calculation, we use the bound state wave
functions as obtained from a Hartree-Fock calculation with a
density-dependent effective interaction. Accordingly,  
in our approach the
spectroscopic factor is the only parameter adjusted to the data.
Concerning the treatment of electron distortion effects the two models
are very different. Whereas in our calculation Coulomb electron
distortion effects are treated to all orders, the CDWBA model of Boffi
{\em et al.} implements electron distortion effects within the high-energy
expansion as briefly mentioned in the theoretical discussion of
section \ref{seclep}. 
In comparing the results obtained with these two
non-relativistic models one can study in how far an exact treatment of
electron distortion effects is required in the analysis of 
(e,e$'$p) reactions.

Our results for the reduced cross sections and the corresponding
spectroscopic factors are also confronted with the  completely relativistic
calculations of Jin {\em et al.} \cite{jin922} and Ud\'{\i}as {\em et
al.} \cite{udi93}.
In line with our approach, the two models handle the electron distortion
in an exact
distorted wave calculation. The main difference with our model occurs
in
 the description 
 of the photoabsorption process and  the
initial and final nuclear system in the (e,e$'$p) process. Jin {\em et
al.}  and  Ud\'{\i}as
{\em et al.}  work in a totally relativistic framework. 
The bound
state wave functions are calculated from the Dirac equation with a
scalar and vector potential which are parametrized fits to
relativistic Hartree potentials. The wave function of the knocked out
nucleon is the solution of the Dirac equation with a relativistic
optical potential. Two different prescriptions for 
the relativistic nuclear current operator are considered. They 
are referred to as the $cc1$ and $cc2$ current operators and
follow
 the
conventions of ref.
\cite{for83}.

\subsection{Parallel kinematics}

In this section we deal with the quasielastic (e,e$'$p) reaction from 
$^{16}$O, $^{40}$Ca, $^{90}$Zr and $^{208}$Pb in parallel kinematics. 
In Table~\ref{kino16} we specify the studied kinematical
conditions. They all correspond with measurements performed at the
NIKHEF electron accelerator.
The (e,e$'$p) cross section for these different target nuclei 
are calculated
in the CDWBA framework as outlined in the previous sections. 
The FSI of the ejected proton with  the
residual nucleus is handled within an OPM. For
the medium-heavy nuclei $^{40}$Ca, $^{90}$Zr and $^{208}$Pb 
the potential as derived from the Schwandt parameterization
\cite{sch82} is considered. This
optical potential is known to provide a good description of the
elastic (p,p$'$) scattering data over a large range of target mass and
incident proton energies. The target nucleus $^{16}$O is out of the
range of nuclei used in the parameterization of this global optical
potential.
Therefore, for the $^{16}$O(e,e$'$p) calculations, we adopt the
optical potential which is directly extracted from a recent analysis
of
elastic
$^{16}$O(p,p$'$) scattering data at $T_p=100$ MeV and use the
parameterization quoted as ''WS'' in ref.~\cite{leu94}.
In order to study the effect of Coulomb distortions the
(e,e$'$p) predictions from the DWBA and CDWBA model are compared. 
We stress that
these two models only differ in the way the Coulomb distortions are
described. In the DWBA they are completely neglected, whereas in the
CDWBA they are treated exactly.

\begin{figure}[htb]
\centering\epsfig{file=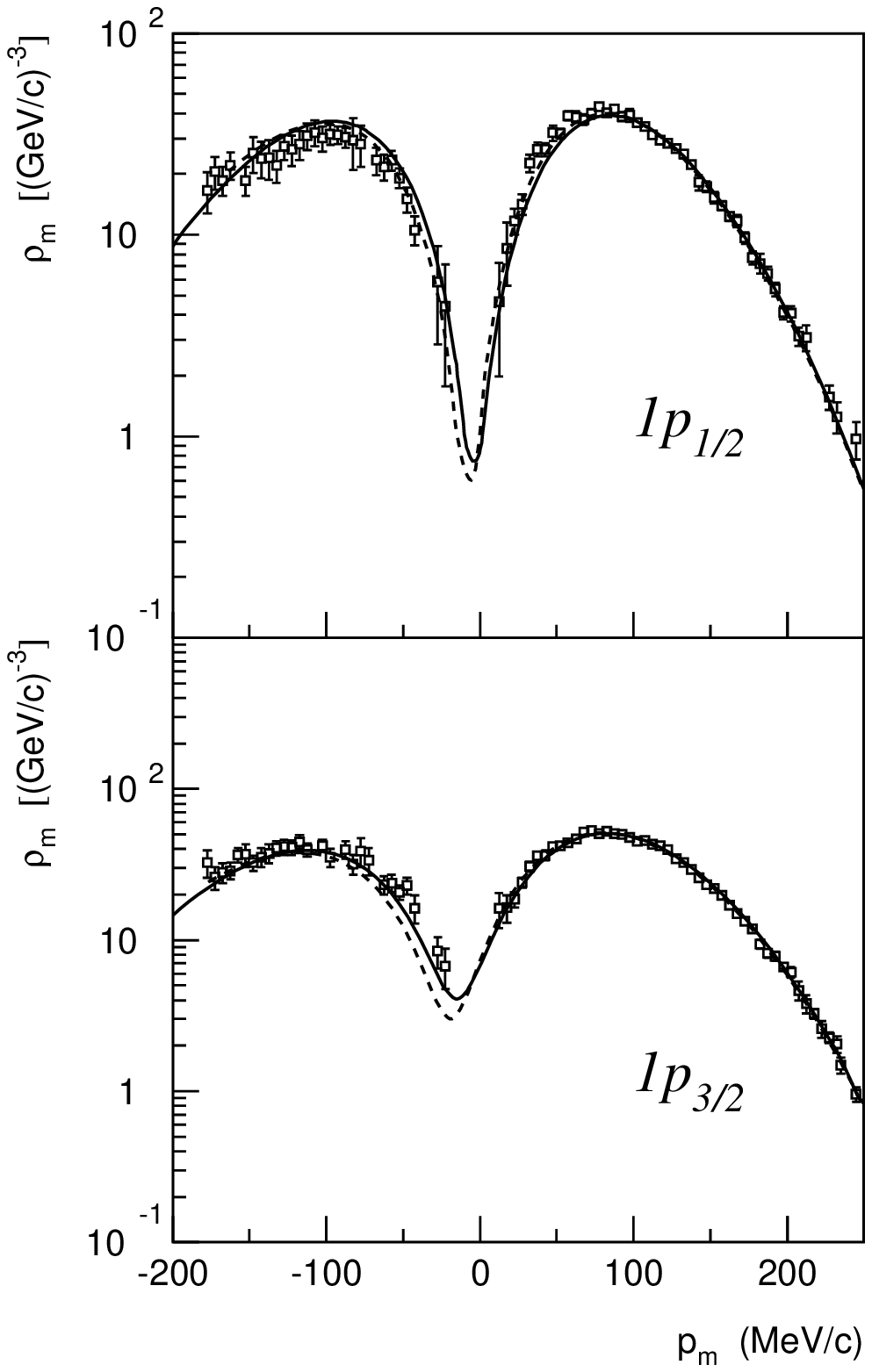,width=11cm}
\caption{Comparison of the DWBA (dashed line) and CDWBA (solid line)
 results for proton knockout %
\mbox{from ${}^{16}$O for} %
  parallel kinematics. The curves are multiplied
with the appropriate spectroscopic factors from Table~\ref{speco16}.
The data are from ref.~\protect\cite{leu94}.
}
\label{o16leuschws.eps}
\end{figure}

The CDWBA reduced cross sections for electro-induced
proton
knockout from the $1p1/2$ and $1p3/2$ shell in ${}^{16}$O 
are confronted with
the NIKHEF data in Fig.~\ref{o16leuschws.eps}.
The DWBA and CDWBA curves for each state are multiplied 
with one and the same spectroscopic factor.
This spectroscopic factor is extracted from a least-square fit of
the CDWBA reduced cross section to the data. 
The multiplication factors as
extracted from our calculation 
adopting the WS optical potential  are
given in Table~\ref{speco16}. 
Table~\ref{speco16} also lists
the spectroscopic factors obtained within 
the non-relativistic 
CDWBA model of the Pavia group \cite{bof93} as reported in
ref.~\cite{leu94}. 
Comparing the results presented in this work and the predictions
outlined in ref.~\cite{leu94},
a similar degree of agreement with the data is reached.
The extracted spectroscopic factors agree within $10\%$.
From
Fig.~\ref{o16leuschws.eps} it is clearly seen that 
the calculated reduced cross
sections
  reproduce the measurements very well 
and electron distortion
effects, although small, improve the agreement with the data
especially for knockout from the $1p3/2$ orbit in $^{16}$O.

\begin{figure}[htb]
\centering\epsfig{file=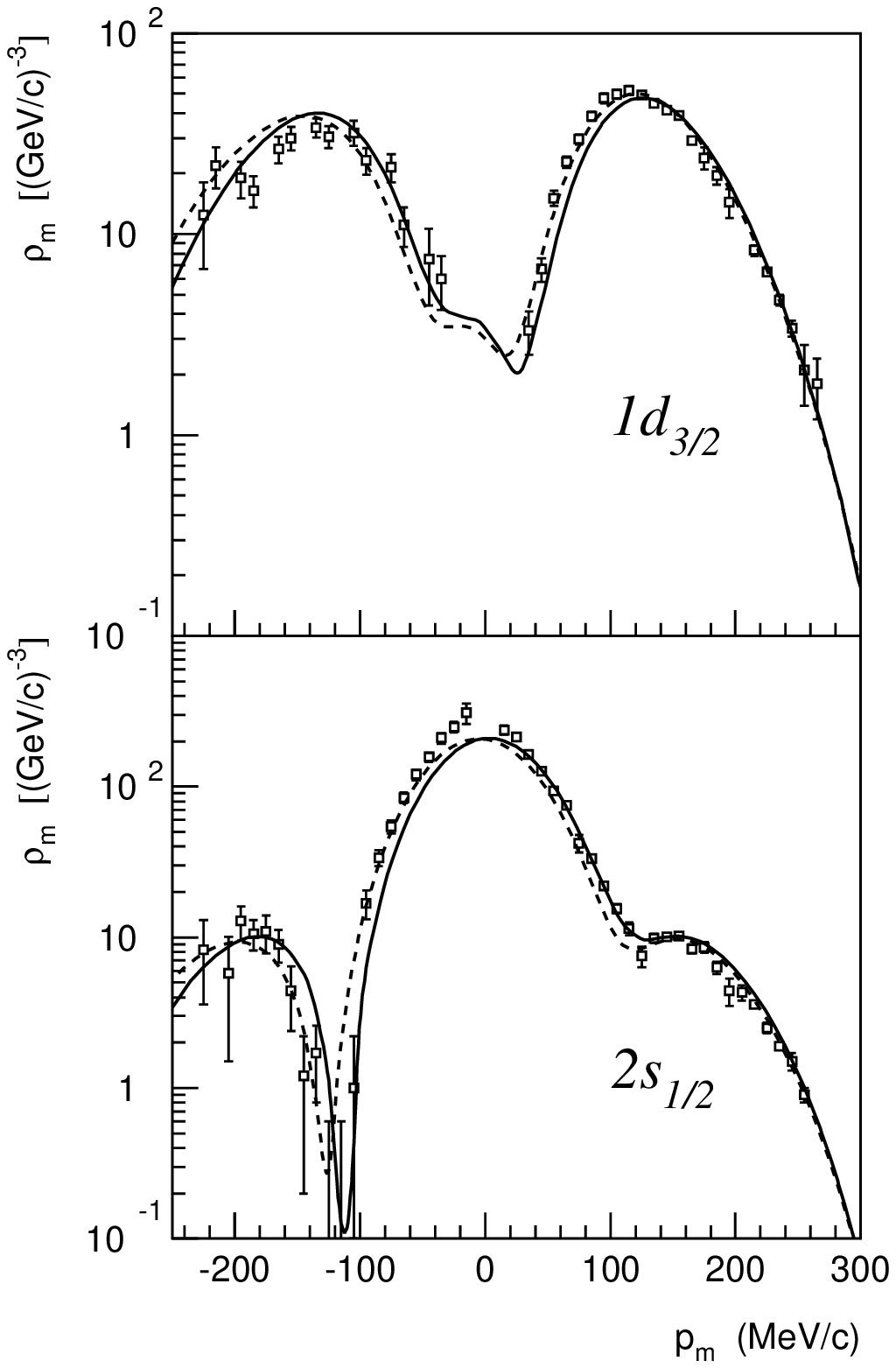,width=11cm}
\caption{Comparison of the DWBA (dashed line) and CDWBA (solid line)
 results for proton knockout
from ${}^{40}$Ca under parallel kinematics. The curves are multiplied
with the appropriate spectroscopic factors (see Table~\ref{specca40}).
The data are  from ref.~\protect\cite{kra90}.
}
\label{ca40kram.eps}
\end{figure}

We also performed calculations for electro-induced
one-proton knockout from the $1d3/2$ and $2s1/2$ shell in $^{40}$Ca.
The results are plotted in Fig.~\ref{ca40kram.eps}. Electron
distortion effects seem to follow the same pattern  as observed for
electron scattering from $^{16}$O, but the effect is now more pronounced.
From the $^{16}$O(e,e$'$p) and $^{40}$Ca(e,e$'$p) results
one can already trace the main 
effects of electron distortion on the reduced cross section in
parallel kinematics: 
\begin{itemize}
\item{Electron distortion
 shifts the reduced cross section towards higher missing
momenta. This can be explained by considering 
that a virtual photon exchanged
between the electron and the nucleus will carry a momentum $\vec{q}^{\rm
\;eff}$ instead of $\vec{q}$ ($q^{\rm eff}>q$). 
From equation~(\ref{keff}) and the
definition of $p_m$  one deduces that
this shift will be decreasing with increasing $p_m$.}
\item{The shape of the reduced cross section 
 is mainly modified at the minima and maxima. Clearly,
electron distortion not only manifests itself in an effective momentum
shift but also in a focusing effect of the electron beam onto the
target nucleus.}
\end{itemize}

The curves in Fig.~\ref{ca40kram.eps}
 are scaled with a spectroscopic factor obtained from a least-square
fit of the CDWBA results to the data. In Table~\ref{specca40} we
compare the spectroscopic factors  from our analysis with
those obtained from the non-relativistic analysis  with the DWEEPY
code~\cite{bof93}
and those extracted from the two 
complete  relativistic calculations by
Jin {\em et al.} \cite{jin922} and Ud\'{\i}as {\em et al.}
\cite{udi93}.

Comparing the spectroscopic factors for proton knockout from the
$1d3/2$ shell two main features can be observed:
\begin{itemize}
\item{The spectroscopic factors obtained with the non-relativistic
models are in very good agreement with each other
but are considerably smaller than the relativistic
values obtained with the $cc2$ nuclear current operator;}
\item{The spectroscopic factors extracted within the relativistic models
seem to be very sensitive to the prescription for the off-shell
 nuclear current
operator. The $cc2$  current operator 
results in a spectroscopic factor for the $2d3/2$ state
that differs with more than $20\%$ from the $cc1$ result. The $cc1$
current operator is obtained from the $cc2$ current operator using the
Gordon decomposition and should produce similar results for on-shell
nucleons.}  
\end{itemize}
The appreciable difference between the relativistic and
non-relativistic approaches is rather surprising considering that the
proton kinetic energies dealt with are typically of the order of $100$ MeV.
According to Jin {\em et al.} \cite{jin942}
and Ud\'{\i}as {\em et al.} \cite{udi93,udi932} the noticeable 
difference between the relativistic and non-relativistic
spectroscopic factors is caused by
the stronger absorptive part
in the relativistic  potentials. Even though all optical
potentials reproduce the elastic proton-nucleus scattering data
to a more or less similar degree, 
the quenching of the  reduced cross section
due to the final-state interaction of the ejected proton with the
residual nucleus
can differ by $15\%$
adopting a relativistic or
non-relativistic  optical potential.
This can be attributed to the behaviour of the optical potential in
the nuclear interior. One could however doubt whether the interior
part of the optical potential can be constrained in elastic proton
scattering processes that are typical surface events.

Hedayati-Poor {\em et al.} \cite{hed95} attribute the difference between the
relativistic and non-relativistic spectroscopic factors to the nuclear
current operator. 
They show that 
the non-relativistic reduction of the relativistic
transition amplitude results in an  effective non-relativistic current
operator which depends on the strong scalar and vector potentials
\cite{wal74}  for the
bound and the continuum single-particle states.
Instead of  using this medium-modified
non-relativistic nuclear current, we adopt the standard non-relativistic
nuclear current operator in our calculations. In our opinion, 
this  is justified as long as
the sensitivity of the relativistic results 
to the choice of the relativistic
nuclear current operator is not cleared up.
   
Concerning the spectroscopic factors obtained for proton knockout from
the $2s1/2$ shell (see Table~\ref{specca40}),
 the different models give very different predictions.
In conformity with the calculation of Ud\'{\i}as {\em et al.}
we describe rather poorly  the  reduced cross section
 around $p_m=0$, especially in the negative missing momentum region.
This  results in a
 spectroscopic factor which is not
very reliable.
However it has to be stressed that, in contrast with what was done in
 the 
analysis of ref. \cite{kra89},
 no attempt has been made to improve
the results by adjusting the parameters of the optical potential 
and/or by adjusting the bound-state wave characteristics 
(rms radii and binding energies).
 
\begin{figure}[htb]
\centering\epsfig{file=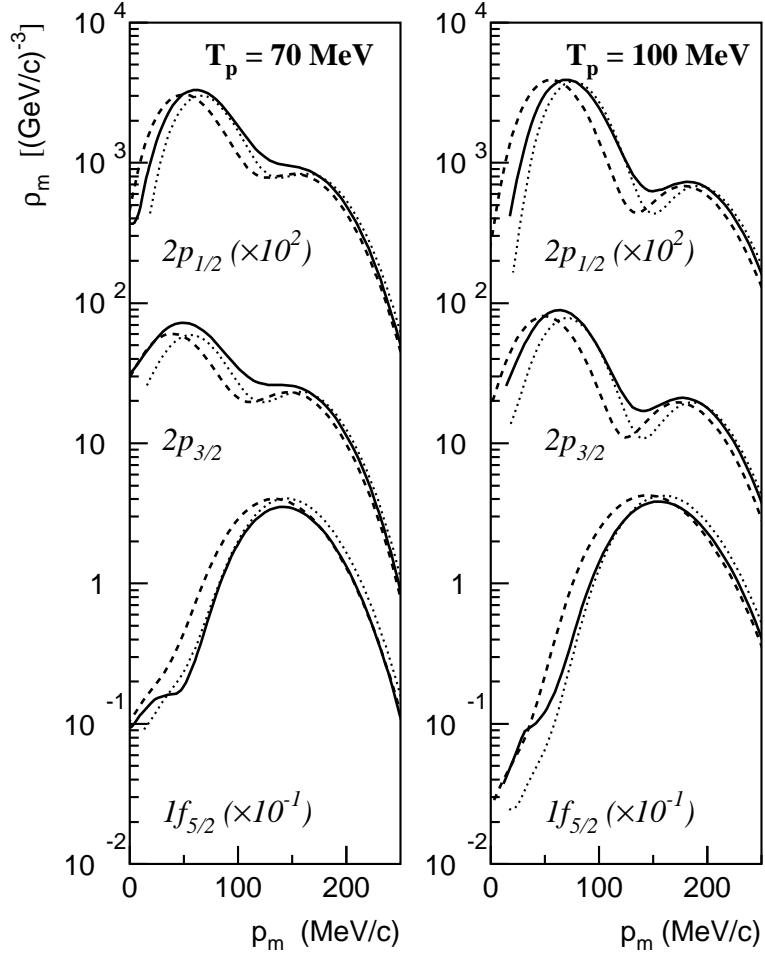,width=12cm}
\caption{Effect of Coulomb distortion on the reduced cross sections
for proton knockout from the three valence shells in ${}^{90}$Zr at
 $T_p = 70$ MeV and $T_p = 100$ MeV.
The dashed line stands for the DWBA result, the dotted line the
EMA result
 and the solid line the complete CDWBA calculation.}
\label{momart.eps}
\end{figure}

The next target nucleus we considered is $^{90}$Zr. The calculations
cover knockout from the different valence shells in $^{90}$Zr for two
different proton kinetic energies ($T_p=70$ and $100$ MeV).
We investigate to what extent the reduced cross sections
for knockout from  the outermost shells (2p1/2, 2p3/2, 1f5/2) 
 are affected by electron Coulomb
distortion effects.
 Secondly the results of the complete calculation
are confronted with the available data. 
In Fig.~\ref{momart.eps} the reduced cross
sections derived  within the
 DWBA (neglecting electron
distortion) and the complete CDWBA framework are compared with the
predictions  adopting the EMA. 

The gross features which were pointed out in the previous sections again
show up.
For the two proton kinetic energies, electron distortion shifts the
reduced cross section towards higher missing momenta. However this
shift is less pronounced in the CDWBA calculation 
than in  the EMA approach.
In
Table~\ref{shiftzr90} we list the missing momenta corresponding with the
first maxima in the reduced cross section 
for knockout from the $2p1/2$ orbit
 for the different steps in the formalism. 
We remark a general behaviour for the two proton kinetic
energies. Including FSI effects which is equivalent with going from
a PWIA to a DWBA approach, a shift towards lower $p_m$ is noticed. 
This shift,
opposite to the shift due to electron distortion, can be
easily explained on the basis of
 an effective proton momentum. The ejected proton
feels an attractive potential (real part of the optical potential)
which causes  the detected proton to have a smaller asymptotic 
 momentum $p_p$
than the momentum $p_p^{\rm eff}$ of the initially struck proton.
Table~\ref{shiftzr90} also shows that this shift towards lower
$p_m$ is increasing with decreasing proton energy.
The latter is easily explained as
the real part of the optical potential induces a shift in the average
measured proton momentum approximately given by \cite{bof93}
\begin{eqnarray}
\vec{p}_p^{\rm \; eff} \approx 
\left(1+\frac{E_p}{p_p^2} <V>\right) \vec{p}_p
\end{eqnarray}
where $<V>$ is the average value of the real part of the optical
potential over the interaction region.

The inclusion of electron distortion effects 
in the model  shifts the reduced cross
section towards higher $p_m$. The shift obtained from the EMA is
larger for $T_p=100$ MeV than for $T_p=70$ MeV since for the latter
the reduced cross section at the peak position corresponds
with a smaller momentum transfer.      
The complete CDWBA calculation produces more than just a shift
towards higher $p_m$. The focusing of the electron beam in the
vicinity of the target nucleus  strongly modifies the maxima and
minima of the reduced cross section with respect to the DWBA
results. Since the extracted
spectroscopic factors are  sensitive to the behaviour of the reduced
cross section at the peaks, an accurate prediction of this focusing
effect is extremely important 
for an accurate deduction of these quantities. 

\begin{figure}[htb]
\centering\epsfig{file=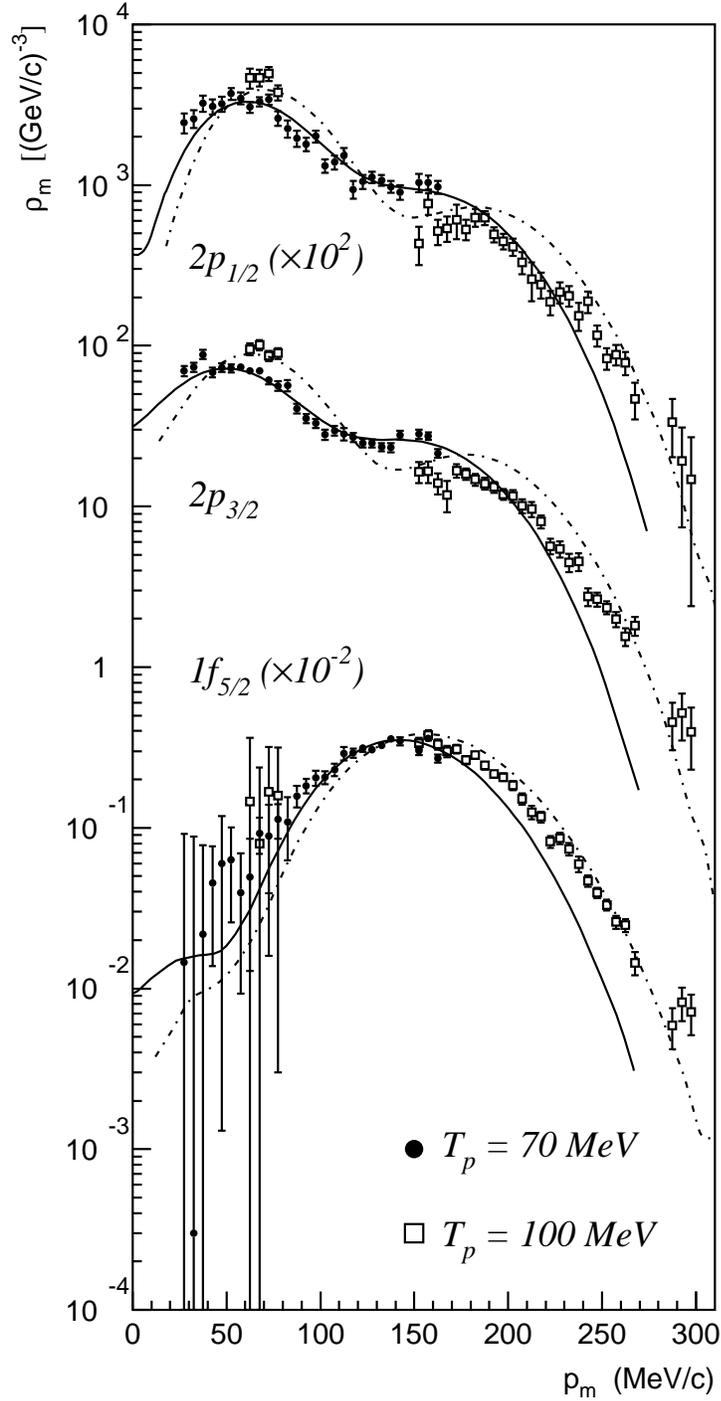,width=11cm}
\caption{Comparison of the CDWBA calculation with the NIKHEF data for
proton knockout from ${}^{90}$Zr (\protect\cite{her87}\protect)
(solid line: $T_p=70$ MeV; dotted-dashed line: $T_p=100$ MeV). The 
 curves are
multiplied with the spectroscopic factors derived for $T_p=70$ MeV
(Table~\ref{speczr90}). }
\label{momallzr.eps}
\end{figure}

In Fig.~\ref{momallzr.eps} the CDWBA results are
confronted with the data.
The different curves are multiplied with the
spectroscopic factors that are determined from  the $T_p=70$ MeV data
(Table~\ref{speczr90}). Firstly, it is clear that for $T_p=70$ MeV
 the calculated  cross sections are
in very good agreement with the data. On the other hand,
 for the $T_p=100$ MeV
data the shape of the measured reduced cross sections is not well
reproduced by the CDWBA calculations
 and, as such, the extracted spectroscopic factors
 can not be considered as reliable.
This conclusion  
agrees with the findings of 
 den Herder in ref.~\cite{her87}.
In ref.~\cite{her87} it was shown that a slight reduction of
the depth of the 
central imaginary part of the optical potential 
resulted in a much better agreement with the data for $T_p=100$ MeV.
With this modified optical potential an equally good fit of the
elastic proton scattering data was obtained. This indicates that low
energy (p,p$'$) reactions are rather insensitive to the depth of the
imaginary part in the nuclear interior. 
The (e,e$'$p) results, however, are
 sensitive to this part of the optical potential.
The second maximum in the
$2p$ reduced cross section reflects the behaviour of the $2p$
single-particle wave function in the nuclear interior. 
Since the overlap is taken with the 
continuum wave functions, the
second maximum is
sensitive to the shape of the continuum wave function in the nuclear
interior.
Given the uncertainties in the optical potential, this behaviour is not
very accurately determined.
As the $1f5/2$
single-particle wave function is more surface peaked,
this also explains  why the $1f5/2$ reduced cross section is not that
sensitive to the depth of the imaginary part of the optical potential.
Clearly, the sensitivity of the reduced cross section to the
parameterization of the
optical potential is a general  weakness of CDWBA models but  
does not affect the
general conclusions with respect to the role of electron distortion on
the reduced cross section.

 The spectroscopic factors
extracted from our $^{90}$Zr(e,e$'$p) calculation
 are systematically larger than the
 values obtained by den Herder \cite{her87}. 
This deviation can be partly attributed to the fact
 that in ref.~\cite{her87} a different optical potential 
  is considered.
Furthermore,
the analysis performed by den Herder  accounts for
 electron distortion effects 
in an approximate way, thus  overestimating the
focusing effect of the electron beam. 

The electro-induced one-proton knockout reaction from $^{208}$Pb is the
ultimate testing case to study electron
distortion effects. The 82 protons in $^{208}$Pb generate a strong 
Coulomb potential felt by the initial and final electron.
We have calculated the $^{208}$Pb(e,e$'$p) reduced cross sections 
for proton emission from 
the $3s1/2$, $2d3/2$, $2d5/2$, $1g7/2$ and $1h11/2$ shells.

\begin{figure}
\centering\epsfig{file=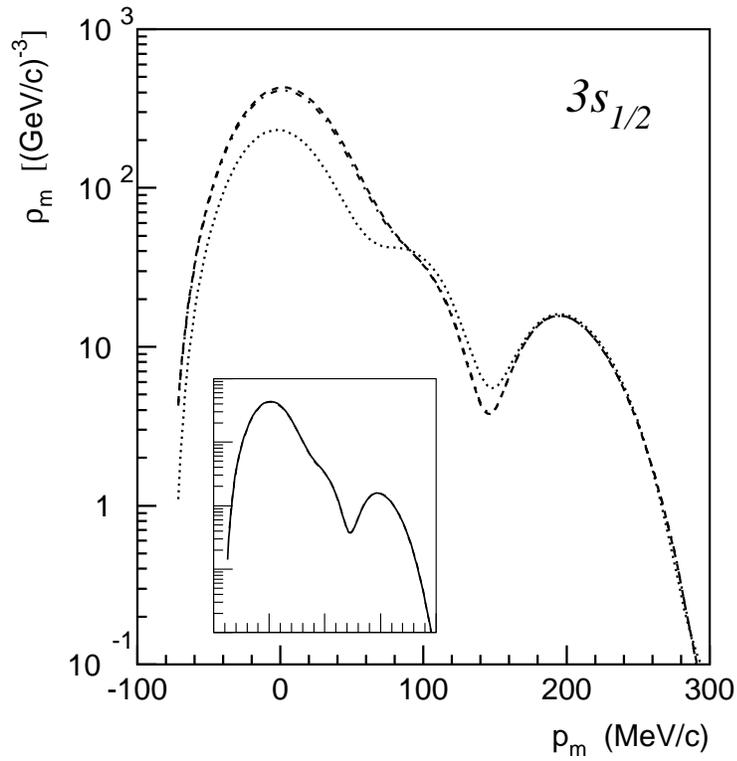,width=11cm}
\caption{
Convergence check of the CDWBA $^{208}$Pb(e,e$'$p) calculation in
parallel kinematics. The electron wave functions are described by
spherical Bessel functions. 
 For the
dotted, dot-dashed and dashed line 
electron partial waves up to $l=30$,$40$ and $50$ are considered.
In the insert
the DWBA calculation (solid line) is compared 
 with the CDWBA calculation when convergence is reached (dashed line).}
\label{converpb208.eps}
\end{figure}

As the effect of Coulomb distortions increases with proton number $Z$,
we consider $^{208}$Pb the ideal target nucleus to illustrate the
numerical accuracy of our technique. The convergence rate for the
electron partial waves is illustrated in
Fig.~\ref{converpb208.eps}.
Convergence is reached for $l=50$ and the code is verified to produce
gradually converging results, which is not evident, considering the
large number of partial waves that has to be considered.
The convergence tests were performed with electron wave functions of
the spherical Bessel type. Accordingly, when convergence is reached
the resulting cross section should coincide with the one obtained in a
DWBA approach, provided that similar model assumptions with respect to
the bound state wave functions and the FSI are adopted. In the insert
of Fig.~\ref{converpb208.eps} it is verified that our CDWBA code bears
this thorough test.

\begin{figure}
\centering\epsfig{file=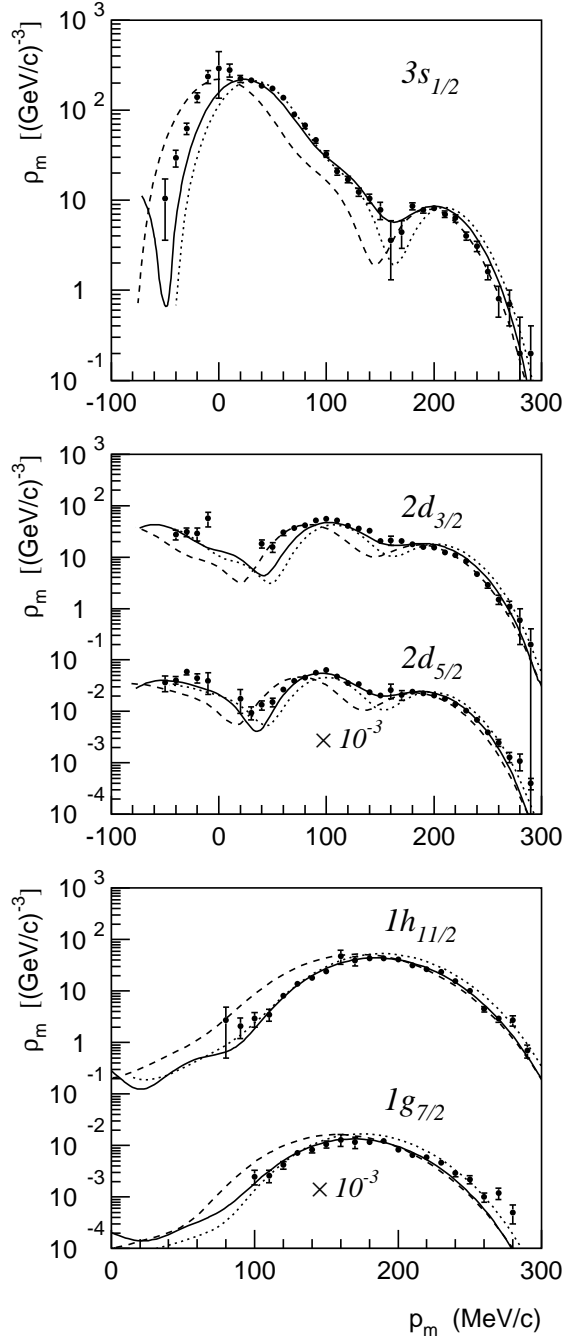,width=9cm}
\caption{The reduced cross sections for electro-induced one-proton
knockout from the valence shells in ${}^{208}$Pb for parallel
kinematics. 
The dashed, dotted
and solid curve give the DWBA, EMA and CDWBA results. The
calculations are compared with the data from
ref.~\protect\cite{qui88}
 and
are multiplied with the spectroscopic factors from
Table.~\protect\ref{specpb208}.
}
\label{momqeffpb208.eps}
\end{figure}

Fig.~\ref{momqeffpb208.eps} shows a comparison of the CDWBA reduced cross
sections with the DWBA and EMA results. All curves are
multiplied with the spectroscopic factors as listed in
Table~\ref{specpb208}.
The spectroscopic factors are derived from a best fit of the CDWBA curve
to the data. The data are 
 well
reproduced in the CDWBA, especially for the positive $p_m$ side. 
We remark that  electron
distortion effects considerably improve on the agreement
with the data. Besides a shift towards higher
$p_m$, the minima and the maxima of the DWBA reduced cross section are
strongly modified when including electron distortion effects in the model.

In Table~\ref{shiftpb208} some characteristics of the first and second
maximum of the $3s1/2$ reduced cross section are listed. 
The numbers clearly demonstrate that 
 the final-state interaction of the ejected nucleon with
the residual nucleus  causes a small shift of the reduced cross section
towards higher $p_m$. 
Electron distortion effects also show up in a shift towards higher $p_m$.
The EMA again overestimates this feature compared with the complete
distorted wave calculation.
Moreover, in the CDWBA model, 
the shift in $p_m$ related to electron distortion 
is more pronounced for the first than for
the second peak. This can be easily understood by considering that
 the two peaks in the $3s1/2$ reduced cross section
correspond with different values for the momentum transfer $q$.
The focusing of the electron beam onto the target nucleus is
reflected
in
an enhancement of the cross section at the peak positions with respect
to the EMA cross section. 
The relativistic calculation by 
Ud\'{\i}as {\em et al.}~\cite{udi93}  predicts a relative 
 enhancement which is
somewhat larger than our estimate. Nevertheless, it is clear that the two
complete distorted wave calculations do not reproduce the strong
focusing effect of electron distortion as observed with the DWEEPY code 
\cite{giu87,giu88}.
The CDWBA model of the Pavia group predicts
an enhancement for the first peak in the 
$3s1/2$ reduced cross section of about  $20\%$
due to the focusing of the electron beam onto the nucleus.  
 This model accounts for electron
distortion effects up to second order in the high-energy approximation.
It has to be stressed that apart from the treatment of electron
distortion effects, the Pavia   and our model are very similar.  
Accordingly, the procedure of treating electron 
distortion effects is the only
plausible explanation for the considerably 
different spectroscopic factors
extracted with the two models. 

Summarizing, the effect of Coulomb distortion in parallel kinematics
can be understood in terms of a shift of the reduced cross section
 towards higher $p_m$
and a small enhancement of the reduced cross section due to the
focusing of the electron beam onto the nucleus.
It is also demonstrated that electron distortion effects become more
important for heavier nuclei  and need
to be treated in a complete distorted wave calculation in order to
extract reliable spectroscopic factors.

\subsection{Constant $\vec{q}-\omega$ kinematics}

\begin{figure}
\centering\epsfig{file=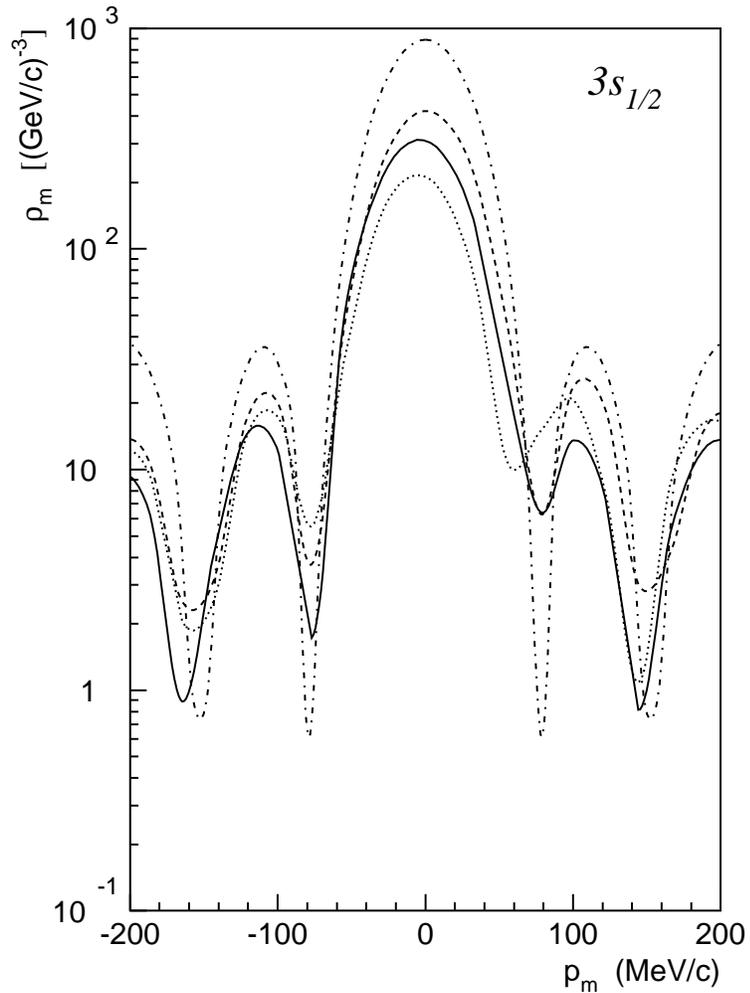,width=11cm}
\caption{The reduced cross section
for proton knockout from the $3s1/2$ shell in ${}^{208}$Pb for
constant $\vec{q}-\omega$ 
 kinematics ($\epsilon=412.3$ MeV, $q=444$ MeV/c,
$\omega=113$ MeV). The dotted-dashed, dashed, dotted
 and solid line
represent the PWIA, DWBA, EMA and CDWBA results. The curves are not
multiplied with a spectroscopic factor.
}
\label{pb208qw.eps}
\end{figure}

We find that for constant $\vec{q}-\omega$ kinematics electron distortion
effects act in a different way than for parallel kinematics.
This is illustrated in Fig.~\ref{pb208qw.eps} where we
investigate electro-induced proton knockout 
from the $3s1/2$ shell in $^{208}$Pb for
quasielastic  kinematics ($p_p\approx q$).
In this calculation, 
the electron energy is the same as the one for  the
parallel kinematics case considered in the previous section.
 The EMA no longer causes
a shift of the reduced cross section towards higher $p_m$
 but now results in a strong quenching of the
reduced cross section around $p_m=0$. 
The peak of the reduced cross section at
$p_m=0 $ in  Fig.~\ref{pb208qw.eps} coincides with the 
reduced cross section for parallel kinematics at $p_m=0$ displayed
in Fig.~\ref{momqeffpb208.eps}.
Moreover, the reduced cross section in parallel
kinematics around $p_m=0$  shows a strong dependence on $q$, i.e.,
 the slightest modification of $q$ considerably affects the value %
 for the reduced cross section. %
Consequently,  a different prescription of $q^{\rm eff}$ than
the one extracted from eq.~(\ref{keff}) can cause a considerable
modification of the EMA
reduced cross section at $p_m=0$ 
for constant $\vec{q}-\omega$ kinematics.

In going from the EMA to the CDWBA,
a strong enhancement of the $3s1/2$ reduced cross section 
 around $p_m=0$ MeV/c is observed. 
From the previous considerations, this
can be ascribed to
a smaller $q^{\rm eff}$-value than
the one adopted in the EMA approach and 
the focusing effect of the
electron beam onto the nucleus.
For constant $\vec{q}-\omega$ kinematics it is therefore more
difficult to disentangle 
the different contributions from electron distortion
as the momentum transfer shift and the focusing effect both might
cause either an
enhancement or quenching of the reduced cross section.

\begin{figure}
\centering\epsfig{file=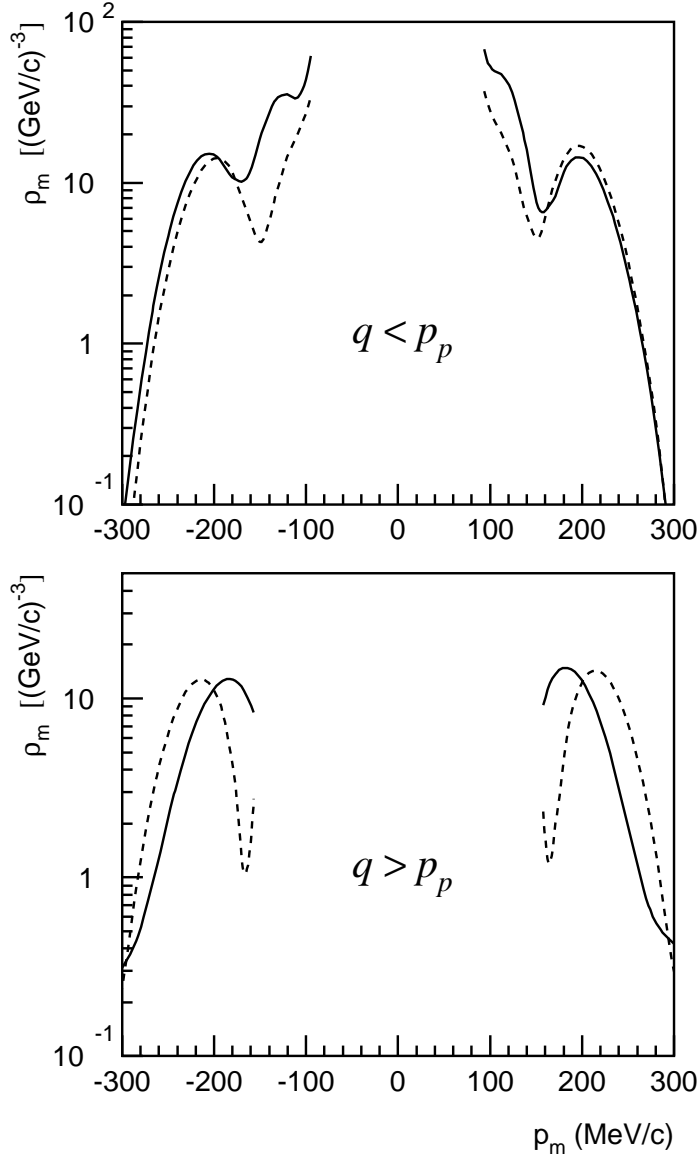,width=11cm}
\caption{The reduced cross section
for proton knockout from the $3s1/2$ shell in ${}^{208}$Pb for
constant $\vec{q}-\omega$ kinematics.
{\em Upper figure}: $\epsilon=412.3$ MeV, $q= 350$ MeV/c,
$\omega=113$ MeV; {\em bottom figure}:
$\epsilon=412.3$ MeV, $ q = 600$ MeV/c,
$\omega=113$ MeV.
The dashed and solid line
represent the DWBA and CDWBA results. The curves are not multiplied
with a spectroscopic factor.}
\label{pb208qwltgt.eps}
\end{figure}

In order to give a complete picture of the role of
 Coulomb distortion 
on the exclusive (e,e$'$p) reaction,
the $^{208}$Pb reduced cross section for constant $\vec{q}-\omega$
kinematics is also studied  for non-quasielastic kinematics.
Two different kinematics are considered, one in the low- ($q>p_p$) 
and one in the
high-energy ($q<p_p$) side of the quasielastic peak.
The results are plotted in Fig.~\ref{pb208qwltgt.eps}.

The 
reduced cross sections show a specific pattern regarding electron
distortion effects. The DWBA
results are shifted
 towards higher (lower) missing momenta for $q<p_p$ ($q>p_p$).
This feature can be easily  explained 
within the EMA.
We stress that
 this EMA shift is only a first estimate of the role of electron
distortion on the exclusive (e,e$'$p)
reduced cross section for non-quasielastic kinematics. It is clear
from Fig.~\ref{pb208qwltgt.eps} that a complete
distorted wave calculation is required to completely account for
electron distortion effects.

\section{Conclusions}

We have analyzed the quasielastic (e,e$'$p)
reaction  from various target nuclei.
A technique is presented to deal with Coulomb distortion effects
in an exact manner keeping the computational time within reasonable limits.
We presented results for
reduced cross sections as a function of missing momentum, 
corresponding
to proton knockout from the outermost shells in ${}^{16}$O,
${}^{40}$Ca, ${}^{90}$Zr and ${}^{208}$Pb considering two different
types of
kinematical arrangements. 

For parallel
kinematics, Coulomb electron distortion causes
a shift of the  cross section towards higher missing momenta. 
This shift
can be partially reproduced by the use of an effective momentum
transfer. Furthermore,the
 focusing effect  of the electron beam onto the nucleus
  mainly affects the maxima and minima
of the reduced cross section. It has been shown that in order to
extract realistic spectroscopic factors an accurate determination of
this effect is highly needed. 

The role of electron distortion in the (e,e$'$p) reaction
for constant $\vec{q}-\omega$  kinematics was  investigated
on the basis of the one-proton knockout reaction
 from the $3s1/2$ shell in $^{208}$Pb.
For quasielastic kinematics the focusing effect and the
effective momentum transfer effect 
caused by Coulomb  electron distortion
are reflected in an enhancement,
respectively, quenching of the DWBA reduced cross section around
$p_m=0$.
For non-quasielastic kinematics, electron distortion effects 
cause a shift of
 the DWBA reduced cross section 
towards higher or lower missing momentum 
depending if we probe the high- or low-energy side of the
quasielastic peak.

\begin{figure}
\centering\epsfig{file=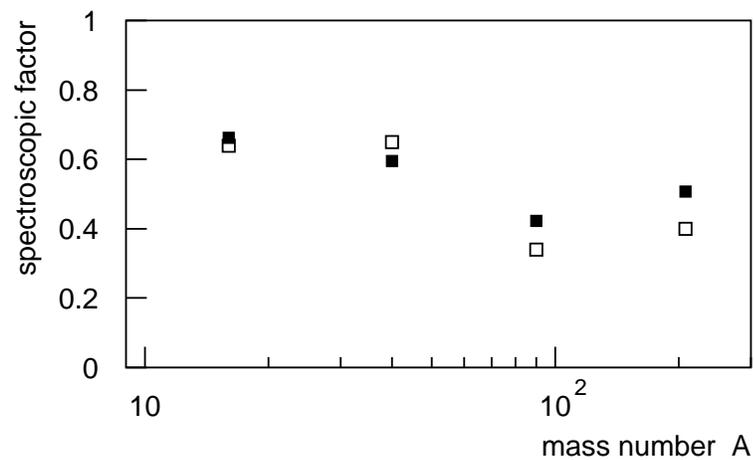,width=12cm}
\caption{Spectroscopic factors derived from the $A$(e,e$'$p) 
reaction to the groundstate of the residual nucleus.
The black squares give the
results within the presented model, whereas the open squares are the
values obtained with the DWEEPY code \protect\cite{bof93} which
 incorporates electron distortion
effects in an approximate manner.}
\label{specplot.eps}
\end{figure}

The spectroscopic factors extracted in our model agree 
 within $20\%$
with the corresponding spectroscopic
factors derived from an analysis applying the CDWBA code of the Pavia
group. As can be seen from Fig.~\ref{specplot.eps}, 
we mostly obtain  larger values. 
This can be partially attributed 
to the way of treating electron distortion
effects. The approximate treatment of electron distortion
by the Pavia group \cite{bof93} tends
 to overestimate the
focusing effect of the electron beam onto the nucleus and produces in
this way
smaller spectroscopic factors than our values.

{\bf Acknowledgment}

This work has been supported by the Fund for
Scientific Research - Flanders (FWO).

\appendix

\section{Feynman amplitude in the CDWBA}
\label{appa}
\begin{figure}
\centering\epsfig{file=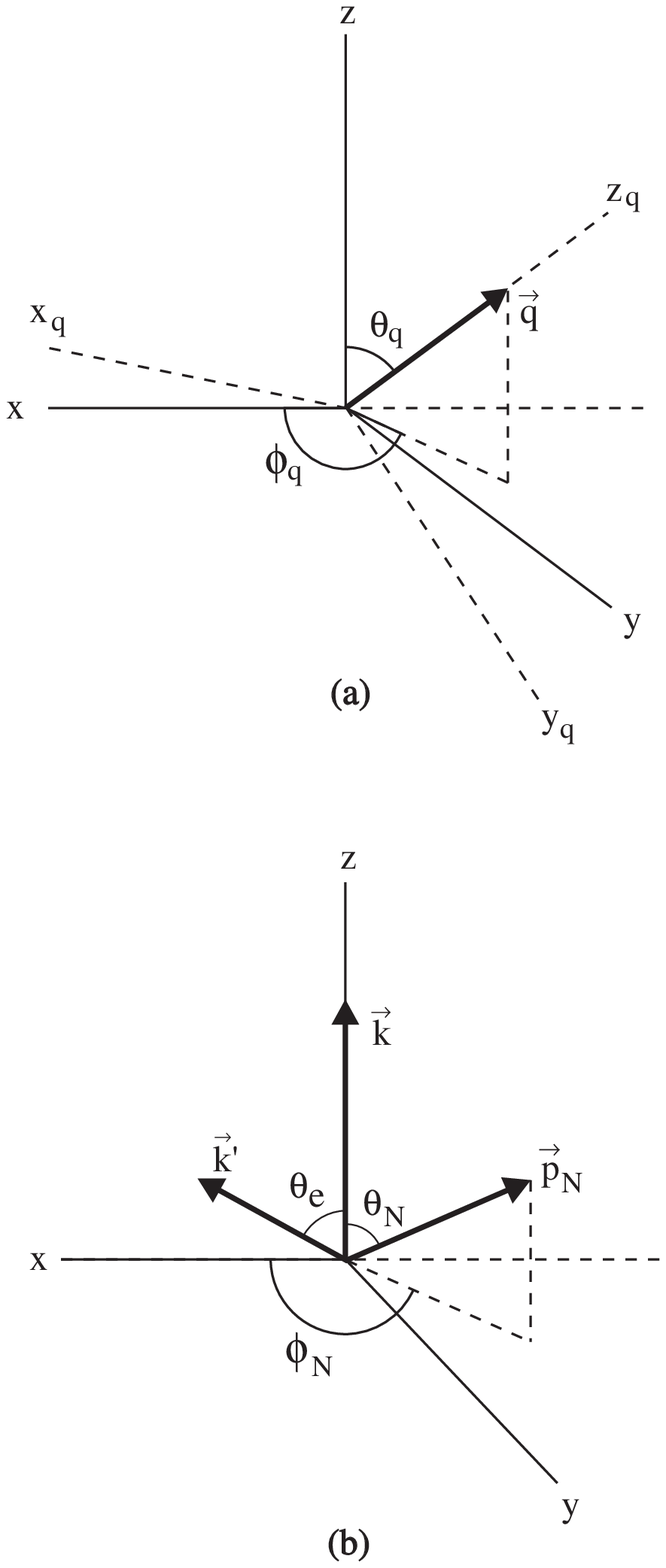,width=8cm}
\caption{Kinematics for the (e,e$'$N) reaction in the CDWBA.}
\label{kincoul2.eps}
\end{figure}

In this appendix we work out the CDWBA feynman amplitude
(\ref{cdwba1}) for the
electro-induced one-nucleon knockout process.
Combining eqs.~(\ref{tranop}) and (\ref{fincdwba}) with
the expressions for the electron charge-current four-vector 
(\ref{elcur}),
 the distorted electron wave functions
(\ref{pardecom})
and the distorted nuclear wave function (\ref{nuclwav}) and
applying some basic properties of the Wigner ${\cal D}$ matrices,
the summation~(\ref{cdwba1}) can be rewritten as
\begin{eqnarray}
\label{totcdwba}
\frac{1}{2}\sum_{m_{s_k} m_{s}}\sum_{m_h m_{s_N}}
\left|
\sum_{LM_{L}} (-1)^{M_{L}}(4\pi)^{3}
 \int_{0}^{\infty}{\rm d}q \;
{\cal A}_{LM_L}( m_{s_k} m_s;m_h m_{s_N};\epsilon\epsilon';q)
 \right| ^2
\end{eqnarray}
with
\begin{eqnarray}
\label{mata}
\lefteqn{{\cal A}_{LM_L}(m_{s_k} m_s;m_h m_{s_N};\epsilon\epsilon';q)
}\nonumber\\
&=&{\cal E}^{{\rm coul}}_{LM_{L}}(m_{s_k}m_s;\epsilon\epsilon';q)
{\cal N}^{{\rm coul}}_{L-M_{L}}(m_h m_{s_N};\omega;q)
\nonumber\\
&-&\frac{q^{2}}{q_{\mu}q^{\mu}+i\eta}
\left\{
{\cal E}^{{\rm mag}}_{LM_{L}}(m_{s_k}m_s;\epsilon\epsilon';q)
{\cal N}^{{\rm mag}}_{L-M_{L}}(m_h m_{s_N};\omega ;q)
\nonumber\right.\\
&&\left.\qquad\qquad
+ {\cal E}^{{\rm el}}_{LM_{L}}(m_{s_k}m_s;\epsilon\epsilon';q)
  {\cal N}^{{\rm el}}_{L-M_{L}}(m_h m_{s_N};\omega ;q)
\right\}\;.
\end{eqnarray}
For the derivation of this expression we have chosen the reference
frame $(x,y,z)$ and the electron and nucleon
scattering angles according to the definitions fixed in
Fig.~\ref{kincoul2.eps}(b).

\vspace{0.5 cm}

\noindent The {\bf electron part ${\cal E}$}  reads as
\begin{eqnarray}
\label{electroncdwba}
&&\left(\begin{array}{c}
{\cal E}^{{\rm coul}}_{LM_{L}}(m_{s_k}m_s;\epsilon\epsilon';q)
\\{\cal E}^{{\rm el}}_{LM_{L}}(m_{s_k}m_s;\epsilon\epsilon';q)
\\{\cal E}^{{\rm mag}}_{LM_{L}}(m_{s_k}m_s;\epsilon\epsilon';q)
\end{array}\right)
= -\frac{e}{\pi}\sum_{j_{1}j_{2}}\sum_{l_{1}l_{2}} 
\left\{e^{i(\delta^{{\rm e},\epsilon'{\rm (tot)}}_{j_{1}}
+\delta^{{\rm e},\epsilon
{\rm (tot)}}_{j_{2}})} i^{-L}
\frac{\hat{l_{2}}}{\sqrt{4\pi}}(-1)^{j_{1}+m_{s_{k}}}\right.\nonumber\\
& &\quad\times< l_{2} 0  1/2 m_{s_{k}}\mid j_{2} m_{s_{k}} >
\frac{(-1)^{L}}{\hat{L}}
< j_{1} -(M_{L}+m_{s_{k}}) j_{2} m_{s_{k}} \mid L
-M_{L}>
\nonumber\\& &\quad\times<l_{1} (M_{L}+m_{s_{k}}-m_{s}) 1/2 m_{s}
\mid j_{1} M_{L}+m_{s_{k}} > {\rm Y}_{l_{1} (M_{L}+m_{s_{k}}-m_{s})} 
(\Omega_{e})\nonumber\\& &\qquad\qquad\qquad\qquad\quad
\left.\times\frac{1}{2}\left(\begin{array}{c}
(1+(-1)^{(l_{1}+l_{2}+L)})
{\cal R}^{{\rm coul}}_{Lj_{1}j_{2}} 
(\epsilon,\epsilon';q)\\
(1+(-1)^{(l_{1}+l_{2}+L)})
{\cal R}^{{\rm el}}_{L j_{1}j_{2}}
(\epsilon,\epsilon';q)\\
(1-(-1)^{(l_{1}+l_{2}+L)}){\cal R}^{{\rm mag}}_{Lj_{1}j_{2}} 
(\epsilon,\epsilon';q)
\end{array}\right)
\right\}\;.
\end{eqnarray} 
The radial integrals ${\cal R}_{Lj_1j_2}$ in this expression are
evaluated in the following way 
\begin{eqnarray}
\label{radint1}
\nonumber\\
\left(\begin{array}{c}{\cal R}^{{\rm coul}}_{L j_{1} j_{2}}
(\epsilon,\epsilon';q)
\\{\cal R}^{{\rm el}}_{L j_{1}j_{2}} (\epsilon,\epsilon';q)\\
{\cal R}^{{\rm mag}}_{L j_{1}j_{2}} 
(\epsilon,\epsilon';q)\end{array}\right)
&=&(-1)^{n-l_{1}}(\bar{l_{2}}-l_{2})(\bar{l_{1}}-l_{1})
\left(
\begin{array}{c}R^{{\rm coul}}_{L j_{1}j_{2}l_{1}l_{2}}
(\epsilon,\epsilon';q)\\
R^{{\rm el}}_{L j_{1}j_{2}l_{1}l_{2}}(\epsilon,\epsilon';q)\\
iR^{{\rm mag}}_{L
j_{1}j_{2}l_{1}l_{2}}(\epsilon,\epsilon';q)\end{array}\right) 
\\\nonumber
\end{eqnarray}
where $n$ stands for $\frac{l_{1}+l_{2}+L}{2}$ in the
 Coulomb and electric radial integrals and for 
$\frac{l_{1}+l_{2}+L-1}{2}$
in the magnetic radial integrals.
It can easily be verified that in case $l_1,l_2$ satisfy $l_1+l_2+L$
even (odd) than $\overline{l}_1,\overline{l}_2$ satisfy 
$\overline{l}_1+\overline{l}_2+L$ even (odd). Moreover, 
$R_{L,j_{1},j_{2}} (\epsilon,\epsilon';q)$ can be evaluated with
either of the
two choices ($l_1,l_2$) or ($\overline{l}_1,\overline{l}_2$) since
for both sets the same result is obtained. The  initial and final radial 
electron partial waves occur in the expressions for the radial integrals 
$R_{L j_1 j_2 l_1 l_2}$ given by
\begin{eqnarray}
\lefteqn{R^{{\rm coul}}_{L j_{1} j_{2} l_{1} l_{2}}(\epsilon,\epsilon';q)
=\int_{0}^{\infty}{\rm d}rj_{L}(qr)}
\nonumber\\
&&\qquad\quad
\times\left[ G^{\epsilon'}_{l_{1}j_{1}}(r)
G^{\epsilon}_{l_{2}j_{2}}(r)<(l_{1}1/2)j_{1}\mid\mid {\rm Y}_{L} \mid\mid
(l_{2}1/2)j_{2}>\right.\nonumber \\  
&&\qquad\quad\quad
+\left. F^{\epsilon'}_{l_{1}j_{1}}(r)
F^{\epsilon}_{l_{2}j_{2}}(r)<(\bar{l_{1}}1/2)j_{1}
\mid\mid {\rm Y}_{L} \mid\mid (\bar{l_{2}}1/2)j_{2}>\right]\;,
\end{eqnarray} 
\begin{eqnarray}
\lefteqn {R^{{\rm mag}}_{L j_{1} j_{2} l_{1} l_{2}}
(\epsilon,\epsilon';q)=i
\int_{0}^{\infty}{\rm d}rj_{L}(qr)}
\nonumber\\
&&\qquad\quad
\times\left[ G^{\epsilon'}_{l_{1}j_{1}}(r)F^{\epsilon}_{l_{2}j_{2}}(r)  
<(l_{1}1/2)j_{1} \mid\mid ({\rm Y}_{L}\otimes \vec{\sigma})_{L}
\mid\mid(\bar{l_{2}} 1/2)j_{2}>\right.\nonumber\\ 
&&\qquad\quad\quad
 -\left. F^{\epsilon'}_{l_{1}j_{1}}(r)G^{\epsilon}_{l_{2}j_{2}}(r)
<(\bar{l_{1}}1/2)j_{1} \mid\mid ({\rm Y}_{L}\otimes \vec{\sigma})_{L}
\mid\mid(l_{2} 1/2)j_{2}>\right] \;,
\end{eqnarray}
\begin{eqnarray}
\lefteqn{R^{{\rm el}}_{L j_{1} j_{2} l_{1} l_{2}}
(\epsilon,\epsilon';q) =}
\nonumber\\
\nonumber\\ &&-\int_{0}^{\infty}{\rm d}r
\frac{\sqrt{L+1}}{\sqrt{2L+1}} j_{L-1}(qr) 
\nonumber\\
&&\qquad\quad\times\left[
G^{\epsilon'}_{l_{1}j_{1}}(r)F^{\epsilon}_{l_{2}j_{2}}(r) <
(l_{1}1/2)j_{1}\mid\mid  ({\rm Y}_{L-1}\otimes \vec{\sigma})_{L} 
\mid\mid
(\bar{l_{2}}1/2)j_{2}> 
\right. 
\nonumber\\ 
&&\left.\quad\qquad\quad
-F^{\epsilon'}_{l_{1}j_{1}}(r)G^{\epsilon}_{l_{2}j_{2}}(r)
<(\bar{l_{1}}1/2)j_{1}\mid
\mid ({\rm Y}_{L-1}\otimes \vec{\sigma})_{L} \mid\mid
(l_{2}1/2)j_{2}>\right]
\nonumber\\&& 
+\int_{0}^{\infty}{\rm d}r
\frac{\sqrt{L}}{\sqrt{2L+1}} j_{L+1}(qr) 
\nonumber\\
&&\qquad\quad\times\left[
G^{\epsilon'}_{l_{1}j_{1}}(r)F^{\epsilon}_{l_{2}j_{2}}(r) <
(l_{1}1/2)j_{1}\mid\mid  ({\rm Y}_{L+1}\otimes \vec{\sigma})_{L} 
\mid\mid
(\bar{l_{2}}1/2)j_{2}> \right. 
\nonumber\\
&& \left.\qquad\quad\quad
-F^{\epsilon'}_{l_{1}j_{1}}(r)G^{\epsilon}_{l_{2}j_{2}}(r)
<(\bar{l_{1}}1/2)j_{1}\mid
\mid ({\rm Y}_{L+1}\otimes \vec{\sigma})_{L} \mid\mid
(l_{2}1/2)j_{2}>\right]\;.
\end{eqnarray}
The following reduced matrixelements are useful in the %
 {\mbox evaluation} %
 of
these integrals
\begin{eqnarray}
&&<(l_1 1/2)j_1 || {\rm Y}_L || (l_2 1/2) j_2 > =
(-1)^{j_1 -\frac{1}{2}} \frac{\hat{j_1}\hat{L}\hat{j_2}}{\sqrt{4\pi}}
\left( 
\begin{array}{ccc}
j_1 & L & j_2 \\ -1/2&0&1/2
\end{array}
\right)
\nonumber\\&&\qquad\qquad\qquad\qquad\qquad\quad
\qquad
\times\frac{ \left(1+(-1)^{l_1+L+l_2}\right)}{2}\;,
\nonumber\\\nonumber\\
&&<(l_1 1/2)j_1 || ({\rm Y}_L\otimes\vec{\sigma})_J || (l_2 1/2) j_2 > =
\hat{j_1}\hat{J}\hat{j_2} 
\left\{
\begin{array}{ccc}
l_1&1/2&j_1\\
l_2&1/2&j_2\\
L&1&J
\end{array}
\right\}
(-1)^{l_1}
\nonumber\\&&\qquad\qquad
\qquad\qquad\qquad\qquad\qquad\quad
\times\frac{\sqrt{6}\hat{l_1}\hat{L}\hat{l_2}}{\sqrt{4\pi}}
\left(
\begin{array}{ccc}
l_1 & L & l_2 \\ 0&0&0
\end{array}
\right)\;.
\end{eqnarray}

\vspace{0.5 cm}

\noindent The {\bf nuclear part ${\cal N}$} in eq.~(\ref{mata}) is
written in terms of the reduced matrixelements ${\cal L}_{{\rm
coul}}$, ${\cal L}_{{\rm el}}$ and ${\cal L}_{{\rm mag}}$ defined 
as follows:
\begin{eqnarray}
\label{redmat}
{\cal L}_{{\rm coul}}(C;q\omega J) &=& 
< 0^+|| M_J^{{\rm n},{\rm coul}}(q) ||(l_h j_h, lj);\omega J > \;,
\nonumber\\
{\cal L}_{{\rm el}}(C;q\omega J) &=& 
< 0^+|| T_J^{{\rm n},{\rm el}}(q) ||(l_h j_h, lj);\omega J > \;,
\nonumber\\
{\cal L}_{{\rm mag}}(C;q\omega J) &=& 
< 0^+|| T_J^{{\rm n},{\rm mag}}(q) ||(l_h j_h, lj);\omega J >\;. 
\end{eqnarray}

We get
 \begin{eqnarray}
\label{nuclearcdwba}
\left(\begin{array}{c}
{\cal N}^{{\rm coul}}_{L-M_{L}}(m_h m_{s_N};\omega ;q)
\\{\cal N}^{{\rm el}}_{L-M_{L}}(m_h m_{s_N};\omega ;q)
\\{\cal N}^{{\rm mag}}_{L-M_{L}}(m_h m_{s_N};\omega ;q)
\end{array}\right)
&=&\sum_{ljmm_{l}} 4\pi i^{-l}\sqrt{\frac{\pi}{2\mu_{N}k_{p}}}
e^{i\delta^{{\rm n},\epsilon_p {\rm (tot)}}_{lj}} 
{\rm Y}_{lm_{l}}(\Omega_{N})\frac{(-1)^{L}}{\hat{L}}
\nonumber\\&&
\times< j_{h}m_{h} j m \mid L-M_{L} >< lm_{l} 1/2m_{s_N} \mid jm >
\nonumber\\&&
\times\left(\begin{array}{c}
{\cal L}_{{\rm coul}}^{\ast}(C;q\omega L)\\ 
-{\cal L}_{{\rm el}}^{\ast}(C;q\omega L)\\
-{\cal L}_{{\rm mag}}^{\ast}(C;q\omega L)
\end{array}\right)\;.
\end{eqnarray}

At this stage we described  the (e,e$'$N) process in its most general form.
All approximations 
with
respect to the
photoabsorption mechanism and 
the final-state interaction (FSI) of the
ejected nucleon with the nucleus are contained 
in the matrixelements ${\cal
L}_{{\rm coul}}(C;q\omega L)$, ${\cal L}_{{\rm
el}}(C;q\omega L)$ and ${\cal L}_{{\rm mag}}(C;q\omega L)$. 
Moreover, besides the fact that we consider the  ultrarelativistic limit, 
electron distortion effects are accounted for exactly.
For the one-body nuclear current operator of the impulse approximation
(\ref{onecur}),
these reduced matrixelements are evaluated in ref.~\cite{ryc88}.

\newpage

\begin{table}
\begin{center}
\begin{tabular}{|l|c|c|c|}
& $\epsilon$ (MeV) & $\omega$ (MeV) & $p_m$ (MeV/c)  \\
\hline
$^{16}$O(e,e$'$p) $^{(a)}$  & 455.8 &  115 & -177 \ldots 265\\
$^{40}$Ca(e,e$'$p) $^{(b)}$ & 460 &   114 & -225 \ldots 285\\
$^{90}$Zr(e,e$'$p) $^{(c)}$ & 346.5 &  81 & 27 \ldots 168\\
$^{90}$Zr(e,e$'$p) $^{(c)}$ & 350.7 &  114 & 62 \ldots 298\\
$^{208}$Pb(e,e$'$p) $^{(d)}$ & 412.3 & 113 & -50 \ldots 300\\
\end{tabular} 
\end{center}
$^{(a)}$ ref.~\cite{leu94}; $^{(b)}$ refs.~\cite{kra89,kra90}
$^{(c)}$ refs.~\cite{her87,her85}; $^{(d)}$ ref.~\cite{qui88}
\caption{Kinematical conditions for the considered
reactions.}
\label{kino16}
\end{table}

\begin{table}[htb]
\begin{center}
\begin{tabular}{|l|c|c|c|}
&$E_x$ (MeV) & CDWBA (this work)  & CDWBA$^{(a)}$   \\
\hline
$1p_{1/2}$ & 0 & {\bf 0.66} & 0.64 \\
$1p_{3/2}$ & 6.3& {\bf 0.54} & 0.51 \\
\end{tabular} 
\end{center}
$^{(a)}$ ref.~\cite{leu94}
\caption{Spectroscopic factors for the ${}^{16}$O(e,e$'$p)${}^{15}$N
reaction.}
\label{speco16}
\end{table}

\begin{table}[htb]
\begin{center}
\begin{tabular}{|l|c|c|c|c|c|}
 &$E_x$ (MeV) &  CDWBA (this work) & CDWBA$^{(a)}$ & rel.
CDWBA$^{(b)}$ & rel. CDWBA$^{(c)}$ \\
\hline
$1d_{3/2}$ & 0 & {\bf 0.60} & 0.65 & 0.80 & 0.76 (0.60)\\
$2s_{1/2}$ & 2.522 & {\bf 0.48} & 0.51 & 0.75 & 0.51(0.44)\\
\end{tabular} 
\end{center}
$^{(a)}$ ref.~\cite{kra89}; $^{(b)}$ ref.~\cite{jin922}; $^{(c)}$
ref.~\cite{udi93} 
\caption{Spectroscopic factors for the ${}^{40}$Ca(e,e$'$p)${}^{39}$K
reaction. The spectroscopic factors
between brackets are obtained with the $cc1$ nuclear current operator
instead of the $cc2$ operator usually
adopted in the relativistic calculations.}
\label{specca40}
\end{table}

\begin{table}[htb]
\begin{center}
\begin{tabular}{|l|c|c|c|c|}
 &$E_x$ (MeV) &  CDWBA (this work)   & CDWBA (this work) & CDWBA$^{(a)}$ \\
 &            &  ($T_p=70$ MeV) & ($T_p=100$ MeV) &  \\
\hline
$2p_{1/2}$ &0& {\bf 0.42} &  0.31 & 0.34\\
$2p_{3/2}$ &1.507 &{\bf 0.51} & 0.36 & 0.44\\
$1f_{5/2}$ &1.745 & {\bf 0.52} & 0.44 & 0.33\\
\end{tabular} 
\end{center}
$^{(a)}$ ref.~\cite{her87}
\caption{Spectroscopic factors for the ${}^{90}$Zr(e,e$'$p)${}^{89}$Y 
reaction. }
\label{speczr90}
\end{table}

\begin{table}[htb]
\begin{center}
\begin{tabular}{|l|c|c|}
 &$p_m$ (MeV/c) & $p_m$ (MeV/c) \\
 &$T_p=70$ MeV & $T_p=100$ MeV \\
\hline
PWIA &61& 61 \\
DWBA &50 & 56 \\
EMA &66 & 77 \\
CDWBA &61 & 66 \\
\end{tabular} 
\end{center}
\caption{The missing momentum corresponding with
 the first peak in the $2p1/2$ reduced
cross section for the different approaches.}
\label{shiftzr90}
\end{table}

\begin{table}[htb]
\begin{center}
\begin{tabular}{|l|c|c|c|c|c|}
 &$E_x$ (MeV) & CDWBA (this work) & CDWBA$^{(a)}$ & rel. CDWBA$^{(b)}$
& rel. CDWBA$^{(c)}$ \\
\hline
$3s_{1/2}$ &0& {\bf 0.51} & 0.40 & 0.71 & 0.70 (0.65)\\
$2d_{3/2}$ &0.35 & {\bf 0.54} & 0.46 & & 0.73 (0.66)\\
$2d_{5/2}$ &1.67 &{\bf 0.41} & 0.39 & & 0.60 \\
$1h_{11/2}$ &1.35 & {\bf 0.43} & 0.42 & & 0.64\\
$1g_{7/2}$ &3.47 & {\bf 0.21}  & 0.19 & & 0.30\\
\end{tabular} 
\end{center}
$^{(a)}$ ref.~\cite{qui88}; $^{(b)}$ ref.~\cite{jin922}; $^{(c)}$
ref.~\cite{udi93,udi96}
\caption{Spectroscopic factors for the 
${}^{208}$Pb(e,e$'$p)${}^{207}$Tl reaction.}
\label{specpb208}
\end{table}

\begin{table}[htb]
\begin{center}
\begin{tabular}{|l||cc||ll|}
 &$p_m$ (MeV/c)& & focusing effect& \\
 & {\em first peak} 
& {\em second peak}& {\em first peak} & {\em second peak}\\
\hline
PWIA &0 &   & 2.04 &      \\
DWBA &3 &195& 1.00 (1.00 ; 1.00) & 1.00 (1.00 ; 1.00) \\
EMA  &32&209& 0.96 & 1.04 \\
CDWBA&24&200& 0.99 (1.08 ; 1.21) & 1.08 (1.14 ; -)\\
\end{tabular} 
\end{center}
\caption{The missing momentum and the value of the reduced cross
section relative to the DWBA result  corresponding with
 the first and second peak of the $3s1/2$ reduced cross section. The
corresponding values obtained by Ud\'{\i}as {\em et
al.}~\protect\cite{udi93} 
and Giusti {\em et
al.}~\protect\cite{giu87,giu88} are listed between brackets.}
\label{shiftpb208}
\end{table}

\newpage

\end{document}